\documentclass[12pt]{article}
\usepackage{amssymb}
\usepackage{amsfonts}
\usepackage{amsmath}
\usepackage{amsmath,amsthm}
\usepackage{graphicx,epsfig, color }

\oddsidemargin 10mm \numberwithin{equation}{section}
\def\be{\begin{equation}}
\def\ee{\end{equation}}
\begin{document}
\begin{center}
{{\bf {  Quantum State Dependence of Thermodynamic Phase
Transition in  4D AdS Gauss-Bonnet Quantum Black Holes Surrounded
With Cloud of Strings}} \vskip 1 cm
 { H. Ghaffarnejad $^{a}$\footnote{E-mail address:
hghafarnejad@semnan.ac.ir}, E. Yaraie $^{a,b}$
        \footnote{E-mail address: eyaraie@semnan.ac.ir}} and M. Farsam $^{a,b}$
        \footnote{E-mail address: mhdfarsam@semnan.ac.ir}}\\
   \vskip 0.5 cm
   \textit{$^a$ Faculty of Physics, Semnan University, P.C. 35131-19111, Semnan,
   Iran}\\
   \textit{$^b$Instituut-Lorentz for Theoretical Physics, ITP, Leiden University, Niels Bohrweg 2, Leiden 2333 CA, The
   Netherlands}\\
    \end{center}
\begin{abstract}
According to the Lovelock theorem where the model \cite{Glav}
could not applicable for Einstein Gauss Bonnet (EGB) gravity in
all 4D curved spacetimes, authors of the reference \cite{ali}
presented an effective model by applying break of diffeomorphism
property. Hence we use the latter model instead of former for
study of thermodynamic behavior of a 4D AdS EGB spherically
symmetric static black hole which surrounded with a cloud of
string. In short our work is extension of the works given by
\cite{Veer,Heg} but not by using \cite{Glav} but by applying
\cite{ali}. Our metric solutions are obtained versus the Hermite
polynomials (quantum harmonic Oscillator) for which eigen values
come from single scale defined by multiplication of the coupling
constants of the model: Namely the regularized GB parameter, AdS
radius, the black hole ADM mass and the string tension. Hence we
claim the obtained metric solution is in fact behavior of
quantized black hole. Because the GB term is originated from
renormalization of quantum matter fields. Also we should pointed
that this kind of quantization is different with the canonical
quantization (Wheeler De Witt).  Our study shows that all phase
transitions of this quantum black hole are dependent to the
Hermite quantum numbers.
\end{abstract}
        \section{Introduction}
In absence of a pure quantum gravity theory which should be valid
for the Planck scale of the spacetime \cite{Fauser}, its semi
classical approach is applicable still. It is known as quantum
field theories propagating on curved spacetimes \cite{Bir,Par}. In
this sense renormalization theory for expectation value of quantum
matter fields stress energy tensor operator, suggests nonsingular
expectation value for it but with an anomaly trace. It is well
known that this conformal anomaly is given by the geometrical
Ricci or Kretschmann scalars or Gauss Bonnet  topological
invariant when quantum matter fields \cite{Bir,Par} are massless.
By regarding these anomaly terms as geometric perspective of
quantum matter field corrections on the Einstein metric equation
which is called as the backreaction equation (see for instance
\cite{Ghaf0} and references therein), one can infer that this
metric backreaction equation or corresponding action functional
can be consider as quantum gravity model in absence of pure
quantum gravity theory which does not still presented. Form this
point of view the EGB gravity which we will use in this work can
be considered as alternative gravity which originates from quantum
matter effects.
 Other approaches such as generalization of number of dimensions of the curved spacetimes to larger than four $D>4$ via
  string theories \cite{Bar,Gas} is also generate such these anomalies \cite{Fio, Nie, Sken}
 .
  Among the various higher order derivative gravitational models which are given in the literature the Lovelock gravity \cite{Lovl}
  is quite special, because it is free of ghost \cite{s1,s2,s3,s6,s7,s8,s9}.
   In fact, there are presented many higher order derivative metric theories which exhibit with Ostrogradsky instability (see \cite{woo, Haya}
   for a good review).
In this sense the actions which contains higher order curvature
terms introduce equations of motions with fourth order or higher
metric derivatives where linear perturbations disclose that the
graviton should be a ghost.
 Fortunately the Lovelock model is free of ghost term which means that this model have field equations involving not more than second
  order derivatives of the metric.
 Action functional of the Lovelock gravity, is given by combinations of various terms as follows.
  The first term is the
cosmological constant $\Lambda,$ the second term is the Ricci
scalar $R=R_\mu^\mu,$ and the third and fourth terms are the
second order Gauss-Bonnet \cite{Lan} and third order Lovelock
terms (see Eq. 22 in ref. \cite{Sus}), respectively. Without the
latter term the Lovelock gravity reduces to a simplest form called
as the Einstein-Gauss-Bonnet (EGB) theory in which the
Einstein-Hilbert action is supplemented with the quadratic
curvature GB term. Importance of this form of the gravity model is
appeared more when we see that it is  generated from effective
Lagrangian of low energy string theory \cite{e1,e2,e3,e4,e5}. In
fact for curved spacetimes with $D>4$ dimensions the Gauss Bonnet
coupling parameter which is calculated by dimensional
regularization method have some regular values but not have for
$D=4$. To resolve this problem the author Glaven and his
collaborator presented a proposal \cite{Glav} but we know now that
their initial proposal does not lead to every well-defined theory
because regularization is guaranteed just some large class of
metrics but not for all metrics. In this sense one can see
\cite{cano, cano1} where its authors explained several
inconsistencies of the original 4DEGB paper by Glavan and Lin.
Particularly, besides pointing out possible problems in defining
the limit or finding an action for the theory, their work also
adds new results to the discussion concerning the ill-definiteness
of second order perturbations even around a Minkowskian background
or the geodesic incompleteness of the spherically-symmetric black
hole geometry presented by Glavan and Lin (See also
\cite{Bayram}). Thus there must be presented other proposals that
can be cover all metric theories.
 Recently a well defined theory is presented \cite{ali} by breaking the diffeomorphism property of the curved spacetime.
 Instead of the former work \cite{Glav} the latter model is in concordance with the Lovelock theorem and thereby, seems more
 to be physical and applicable. For instance FRW cosmology of the latter model is studied in \cite{ali2} and showed success of the model versus
 \cite{Glav}.
     \\
Despite that the gravity model \cite{Glav} is sick in 4
dimensional curved space time but idea of authors of \cite{Veer}
is still useful in which  by adding action of string clouds with
the model \cite{Glav} they obtained metric solution of a charged
4DEGB black hole. Then they are studied thermodynamic properties
of the obtained black hole metric solution. By looking at their
work one can infer that the string tension changes all
thermodynamic variables of the black hole except its  entropy. By
studying the heat capacity behavior they obtained that the smaller
black holes are locally stable. Due to the surrounding cloud of
strings, they obtained a black hole phase transition where a large
unstable black holes transit to a globally thermodynamically small
stable black holes with negative free energy. Their results
demonstrate that the Hawking`s evaporation leads to a
thermodynamically stable remnant with no temperature. We should
point that authors of the work \cite{Veer} did not considered the
cosmological constant effect as vacuum de Sitter or Anti de Sitter
space.
\\
On the other side we should point to idea which is used in \cite{Heg} by applying \cite{Glav} as follows. Authors of the work \cite{Heg}
studied thermodynamics of the 4DAdSEGB black hole in the extended phase space where the cosmological constant behaves as pressure of the AdS
vacuum space. They obtained that the black hole exhibits with a phase transition similar to that of van der Waals system. They also showed
that this black hole exhibited with the Joule-Thomson expansion. They do not used  effects of surrounding clouds of string tension in their work.\\
As an extension of the works \cite{Veer, Heg} we like to have
study effects of string clouds on thermodynamic behavior of
4DAdSEGB chargeless black hole by applying the gravity model
\cite{ali} not the sick model \cite{Glav}.
\\
There are many published works in the literature where the higher
order derivative gravities such as Lovelock or its reduced forms
are applied to study black hole thermodynamics (see for instance
\cite{C1,C2,C3,C4,C5,C6,C7,C8,C9,C10,C11,C12,C13,C14,C15,C16,C17,C18,C19,C20,C21,C22,C23,C24,C25,C26,C28,C29}).
The Lovelock gravity or its reduced models are studied in
cosmological approaches too. In this sense we like to point to
some published works such as follows. For instance authors of the
work \cite{Vas1} have assumed that the graviton is massless or
nearly massless motivated by results of $GW170817$.
 This indicated that speed of the gravitational waves is equal to that of light's and thus the fundamental carrier
 of the gravitational force is nearly massless. The gravitational effects is not short range interaction. Specifically by using an EGB extension
 of Horndeski theories they showed that some classes of the latter theories may be revived and become viable in view of the  $GW170817$ result.
  In ref. \cite{Vas2} they showed how the normal EGB theories can be viable in view of $GW170817$. As an extension of non-minimally coupled scalar
  field to Ricci scalar theory in the context of EGB model one can see \cite{Vas3} and seek what is the swampland criteria in the context of a massless
  graviton EGB theory? To do so \cite{Vas4,Vas5} should be followed. Also authors of the work \cite{chen} checked the validity of the weak cosmic
  censorship conjecture for the 4D charged EGB black hole and considered the effect of the GB coupling constant on the validity of the weak cosmic
  censorship conjecture.
  Hawking studied in a new attempt in a fascinating way, AdS black holes phase transition at a first time which has been called as extended phase space
   where the negative cosmological constant behaves as pressure of AdS vacuum space acts on the local black hole horizon .
\cite{Hawking}.
  The extended phase space represents a phase space in which the traditional first law of black hole thermodynamics is corrected by an additional
  $VdP$ work so that the cosmological constant is regarded as  thermodynamic pressure of AdS space which affects on the black hole.
   In this sense its conjugate variable is thermodynamic volume.
  A large number of investigations have been done on this concept during last years and one can follow them for instance at these works:
\cite{Kast,Dolan,Kubiz1,Dutta,Gu,Zhang1,Cai2,Zhao1,Mo2,Altam,Moliu,Kubiz2,Zou,Zhao2,Ghaf1,Ayd,Hans,Ghaf2,Ghaf3,Ghaf4,feng,fu,Kyang,Almen}.
Particularly for Van der Waals behavior of AdS black holes which
can be can followed some published works such as follows
\cite{Bib1,Bib2,Bib3,Bib4,Bib5}. As a different model one can see
\cite{Juan} where the authors obtained a small to large black hole
phase transition by generalizing  the well known Einstein Maxwell
Dilaton gravity with an additional vector field supporting
spherical or hyperbolic horizons. In this view the strings fluids
as gravitational matter source which has received less attention
in solving the gravitational equations and we like consider in
this work.
 As an important role of the string fluids at cosmology it is enough to say that the rapid expansion of the universe
  during inflation is thought to be related to the expansion of cosmic strings.
\cite{Strom,Let,Rich,Yadav,Gan,Bronn}. In this paper we would like
to investigate extended thermodynamics  phase transition of
4DAdSGB black hole metric solution obtained from \cite{ali} in
presence of strings fluid. We will show important and critical
role of the used string tension which how can  present
quantization condition on the event horizon of the obtained black
hole metric solution. Setup of this
work is as follows.\\
In section 2, we review 4DAdSGB gravity \cite{ali} in presence of
strings fluid. Then in sections 3  and 4 we solve metric solution
of a spherically symmetric static curved space time with a black
hole topology. Our obtained metric solutions are described  by the
well known Hermite polynomials. In section 5 we investigate
various aspect of the extended thermodynamics of 4DAdSGB black
hole in presence of strings fluid for ground state and first
excited state. To do so we study $P-V$ criticality and probe
global stability of the black hole system by plotting diagram of
the Gibbs free energy versus the temperature where there is seen a
swallowtail behavior (coexistence of phases) for pressures higher
than the critical pressure (see figures 3 and 8). Then we
investigate possibility of small to large black hole phase
transition and also Hawking-Page phase transition. Last section
denotes to concluding remark and outlook.
 ~
\section{$4D$ AdS GB gravity with strings
fluid} According to the work  \cite{ali} we know a constraint EGB
gravity in $D\to4$ limit which is given by the first term in the
following action functional. $I_{\mathrm{matter}}$ Second matter
term in the following action is assumed to be clouds of strings
fluid.
\begin{equation}\label{action}I=\frac{1}{16\pi G}\int
dtd^3xN\sqrt{\gamma}\mathcal{L}_{EGB}^{4D}+I_{\mathrm{matter}},
 \end{equation} with
\begin{equation}
 \mathcal{L}_{\mathrm{EGB}}^{4D}=
 2R-2\Lambda-\mathcal{M}+\frac{\tilde{\alpha}}{2}[8R^2-4R\mathcal{M}-\mathcal{M}^2-\frac{8}{3}(8R_{ij}R^{ij}-4R_{ij}\mathcal{M}^{ij}
 -\mathcal{M}_{ij}\mathcal{M}^{ij})],
\end{equation}
where $G$ is the Newton`s gravitational coupling constant,$R$ and
$R_{ij}$ are  the Ricci scalar and the Ricci tensor of the spatial
metric $\gamma_{ij}$ respectively and
\begin{equation}\mathcal{M}_{ij}=R_{ij}+\mathcal{K}_k^k\mathcal{K}_{ij}-\mathcal{K}_{ik}\mathcal{K}^{k}_j,~~~~\mathcal{M}=\mathcal{M}_i^i
\end{equation}
with
\begin{equation}\mathcal{K}_{ij}=\frac{1}{2N}(\dot{\gamma}_{ij}-2D_iN_j-2D_jN_i-\gamma_{ij}D_kD^k\lambda_{GF}).\end{equation}
Here a dot denotes time derivative $t$ and all the effects of the
constraint stemming from the gauge-fixing (GF) are now encoded in
Lagrange multiplier $\lambda{GF}$.  $D_i$ is spatial covariant
derivative and re-scaled regular EGB coupling constant
$\tilde{\alpha}$ is defined versus the irregular GB coupling
constant itself $\alpha_{GB}$ in $D\to4 $ dimensions limit as
$\tilde{\alpha}=(D-4)\alpha_{GB}.$ The above EGB gravity action
satisfies the following gauge condition for all spherically
symmetric and cosmological backgrounds (see \cite{ali} and
\cite{ali2}).
\begin{equation}\sqrt{\gamma}D_kD^k(\pi^{ij}\gamma_{ij}/\sqrt{\gamma})\approx0\end{equation}  In
fact the above EGB action is generated from ADM decomposition of
the 4D background metric as $1+3$ dimension as follows.
\begin{equation}\label{met1}ds^2=g_{\mu\nu}dx^\mu dx^\nu=-N^2dt^2+\gamma_{ij}(dx^i+N^idt)(dx^j+N^jdt)\end{equation}
where $N, N_i, \gamma_{ij}$ are the lapse function, the shift
vector, and the spatial metric respectively. $\gamma$ factor in
the action (\ref{action}) is absolute value of determinant of the
spatial metric $\gamma_{ij}.$ This ADM decomposition is done on
the background metric to remove divergent boundary term of the
higher order metric derivative in the GB term of the action
functional (\ref{action}) in general 4D form \cite{ali}. First
term in the  theory  defined  by (\ref{action}) has the time
re-parametrization symmetry $t\to t=t(t^\prime).$ We now set the
matter source $I_{\mathrm{matter}}$ to be
 the Nambu-Goto action \cite{Let} (see also page 100 in ref. \cite{Bar}) which explains the dynamics of
relativistic strings as follows.
\begin{equation}\label{NG}
 I_{\mathrm{NG}}=\int_{\Sigma} \rho\sqrt{\mathfrak{g}} d\sigma^{0} d\sigma^{1}
\end{equation}
where  $\rho$ is tension (mass per unit length or linear mass
density) in the string and worldsheet of strings can be
parameterized by local coordinates $(\sigma^{0}, \sigma^{1})$.
$\mathfrak{g}$ is absolute value of determinant of induced metric
$\mathfrak{g}_{ab}$ and the bivector $\Sigma^{\mu\nu}$ related to
strings worldsheet are given respectively by
\begin{equation}\label{met3}
\mathfrak{g}_{a b} = g_{\mu \nu} \frac{\partial x^{\mu}}{\partial
\sigma^{a}} \frac{\partial x^{\nu}}{\partial \sigma^{b}}
\end{equation}
and \begin{equation} \label{bi}\Sigma^{\mu \nu}=\epsilon^{ab}
\frac{\partial x^{\mu}}{\partial\sigma^{a}} \frac{\partial
x^{\nu}}{\partial \sigma^{b}},
\end{equation}
where $\epsilon^{ab}$ is two dimensional Levi-Civita tensor
density $\epsilon^{01}=-\epsilon^{10}=1$. Strings fluid is
described by the energy momentum tensor
\begin{equation}\label{stress}
T^{\mu \nu}=\mathfrak{g}^{-\frac{1}{2}}\rho \Sigma^{\mu
\delta}\Sigma_{\delta}^{\nu}=\frac{2\partial\mathfrak{L}_{NG}}{\partial
g^{\mu\nu}}.
\end{equation}
We now investigate to solve the metric equation of the above model
for a spherically symmetric static black hole space time.
\section{4D AdS GBBH surrounded by
strings cloud} By comparing the metric line element (\ref{met1})
with general form of a spherically symmetric state 4D metric line
element
\begin{equation}\label{met}
ds^2=f(r)dt^2-f(r)^{-1}dr^2-r^{2}d^{2}\Omega
\end{equation}
we infer that the lapse function, and the shift vector, the
spatial metric components and lagrange multiplier should be $r$
dependent so that we can write
 \begin{equation} N=f(r)=\frac{1}{\gamma_{rr}},~~~N_i=0,~~~\gamma_{\theta\theta}=r^2,~
 ~~\gamma_{\varphi\varphi}=r^2\sin^2\theta,~~~\lambda_{GF}=\lambda_{GF}(r)\end{equation} where $$\mathcal{L}^{4D}_{EGB}=
 -2\Lambda-\frac{3q^{\prime2}}{2r^4f}-\frac{2(rf^\prime+f-1)}{r^2}$$\begin{equation}\label{EGB}+\tilde{\alpha}\bigg\{
 +\frac{4(f-1)f^\prime}{r^3}-\frac{2(f-1)^2}{r^4}+\frac{q^{\prime2}(rf^\prime+f-1)}{r^6f}-\frac{q^{\prime4
 }}{8r^8f^2}\bigg\}\end{equation} with
 \begin{equation}\label{Q}q=r^2\sqrt{f}\lambda^\prime_{GF}\end{equation} in which $\prime$ denotes to derivative with respect to $r$.
 To calculate explicit form of
 the string action functional (\ref{NG}) for the spherically symmetric static
 background metric (\ref{met}) we should use a static gauge \cite{Bar} where the commoving time $\sigma^0$ in worldsheet $x^{\mu}(\sigma^0,\sigma^1
 )$ is equal to the time $t$ as
  $\sigma^0=t=t_0=constant$ and we should assume the static string is along to the radial direction $r$ of worldsheet $x^{\mu}(\sigma^0,\sigma^1)$
  such that
  \begin{equation}\label{radstring}r=r(\sigma^0,\sigma^1)=F(\sigma^1)
  ,~~~t=\sigma^0,~~~\theta(\sigma^0,\sigma^1)=\varphi(\sigma^0,\sigma^1)=0.\end{equation} Here
   we choose an open string which one edge of the world sheet to be the curve $\sigma^1=0$ and the other edge to be the curve $\sigma^1=a$
   such that $\sigma^1\in[0,a]$ for an open string with arbitrary shape $F(\sigma^1).$ In this case non-vanishing components of the induced metric
   (\ref{met3})
   reads \begin{equation}\label{met4} \mathfrak{g}_{tt}=f(r),~~
   ~~~\mathfrak{g}_{rr}=\frac{-1}{f(r)}\bigg(\frac{dF(\sigma^1)}{d\sigma^1}\bigg)^2, ~~~\sqrt{\mathfrak{g}}=\sqrt{|det(\mathfrak{g})|}=\frac{dF(\sigma^1)
   }{d\sigma^1}.
   \end{equation}
   Spherically symmetric property of the background metric (\ref{met}) causes that the string tension $\rho$ and bivector $\Sigma^{\mu\nu}$
   to be dependent to $r$ alone. Covariant conservation of the string stress tensor (\ref{stress}) leads to the condition
   $\nabla_{\mu}(\rho\Sigma^{\mu\nu})=0$ (see ref. \cite{Let}) which for the metric equation (\ref{met}) we will
   have stress energy tensor components of the Numbu Goto (NG) relativistic string as follows.
   \begin{equation}\label{bivec}T^t_t=T^r_r=-\rho\Sigma^{tr}=\frac{C}{r^2}\end{equation}
   where dimensionless parameter $C$ is a integral constant.
   By regarding the stress energy tensor definition $T^{\mu\nu}=\frac{2\partial \mathfrak{L}_{NG}}{\partial
   g^{\mu\nu}}$ given by (\ref{stress}) and by substituting (\ref{bivec}) and metric
   components (\ref{met}) we can obtain explicit form of the NG string Lagrangian density as follows.
   \begin{equation}\label{NGG}\mathfrak{L}_{NG}=\sqrt{\mathfrak{g}}\rho=\frac{C}{4r^2f^2}.\end{equation}
   By substituting  (\ref{EGB}) and by integrating first term of the action functional (\ref{action}) on the 2-sphere
    $0\leq\theta\leq\pi$, $0\leq\varphi\leq2\pi$
   we obtain $I_{EGB}=\int dtdr\mathfrak{L}$ for the first term of the action functional (\ref{action}) where $\mathfrak{L}$ is
 defined by \begin{equation}\label{4G}\mathfrak{L}=\frac{r^2f}{4G}\mathcal{L}_{EGB}^{4D}.\end{equation} Now we are in position
 to obtain total lagrangian density of the system by
 adding (\ref{NGG}) and (\ref{4G}) as follows.
$$\mathfrak{L}_{total}=\frac{r^2f}{4G}\bigg\{\frac{6}{\ell^2}-\frac{3q^{\prime2}}{2r^4f}-\frac{2(rf^\prime+f-1)}{r^2}+\tilde{\alpha}\bigg\{
 \frac{4(f-1)f^\prime}{r^3}-\frac{2(f-1)^2}{r^4}$$\begin{equation}\label{tot}+\frac{q^{\prime2}(rf^\prime+f-1)}{r^6f}-\frac{q^{\prime4
 }}{8r^8f^2}\bigg\}\bigg\}+\frac{C}{4r^2f^2}\end{equation} where we substituted identity between the cosmological constant and radius of 4D AdS space
 as $\Lambda=-\frac{3}{\ell^2}.$ By substituting \begin{equation}\label{qQ}q^{\prime2}=Q^{\prime2}+\xi\end{equation} in which
 \begin{equation}\label{xi}
 \xi=4r^2f(rf^\prime+f-1)-\frac{6r^4f}{\tilde{\alpha}}
 \end{equation} the total Lagrangian density (\ref{tot}) reads
$$\mathfrak{L}_{total}=\frac{r^2f}{4G}\bigg\{\frac{6}{\ell^2}-\frac{2(rf^\prime+f-1)}{r^2}+\tilde{\alpha}\bigg\{
\frac{4(f-1)f^\prime}{r^3}-\frac{2(f-1)^2}{r^4}\bigg\}\bigg\}$$\begin{equation}\label{tot1}\frac{r^2f}{4G}\bigg\{\xi\bigg[\frac{\tilde{\alpha}
(rf^\prime+f-1)}{r^6f}
-\frac{3}{2r^4f}\bigg]-\frac{\tilde{\alpha}(Q^{\prime4}+\xi^2)}{8r^8f^2}\bigg\}+\frac{C}{4r^2f^2}.\end{equation}
It is easy to check that Euler Lagrange equation for the field $Q$
reduces to the following conservation equation.
 \begin{equation}\frac{\partial\mathfrak{L}_{total}}{\partial Q^\prime}=-\frac{\tilde{\alpha}Q^{\prime3}}{8Gfr^6}=constant\equiv D^3\end{equation}
 for which we have \begin{equation}\label{q}Q^{\prime}=-2Dr^2\bigg(\frac{Gf}{\tilde{\alpha}}
 \bigg)^\frac{1}{3}.\end{equation} This conservation equation helps us to obtain explicit form of the undetermined gauge fixing Lagrange multiplier as
 follows.
 \begin{equation}\lambda_{GF}(r)=\int^{r}\frac{dr_1}{r_1^2\sqrt{f(r_1)}}\times\end{equation}$$\bigg[\int^{r_1}dr_2
 \sqrt{\bigg(\frac{8GD^3r_2^6}{\tilde{\alpha}f(r_2)}\bigg)^\frac{2}{3}+r_3^2f(r_2)f^\prime(r_2)+r_2^2f^2(r_2)-r^2_2f(r_2)-
 \frac{6r_2^4f(r_2)}{\tilde{\alpha}}}\bigg]$$ where we substituted (\ref{qQ}), (\ref{xi})
 and (\ref{q})  into the relation (\ref{Q}).
By substituting (\ref{xi}) and (\ref{q}),  the total Lagrangian
density (\ref{tot1}) reduces to the following
form.\begin{equation}
 \mathfrak{L}_{total}=\frac{\tilde{\alpha}ff^{\prime2}}{2G}+\frac{2}{G}\bigg(\frac{\tilde{\alpha}(f-1)}{r}-r\bigg)
 ff^{\prime}-\frac{2f(f-1)}{G}
  \end{equation}
 $$+\frac{3fr^2}{2G}\bigg(\frac{1}{\ell^2}+\frac{3}{4\tilde{\alpha}}\bigg)-\frac{D^4r^2}{2}\bigg(\frac{Gf}{\tilde{\alpha}}\bigg)^\frac{1}{3}
 +\frac{C}{4r^2f^2}$$ for which one can obtain Euler Lagrange equation for the metric potential $f(r)$ as follows.
 \begin{equation}\label{euler}2f^4f^{\prime\prime}+f^3f^{\prime2}+\frac{4f^3(f-1)}{\tilde{\alpha}}-
 \frac{4f^4(f-1)}{r^2}\end{equation}
 $$-\frac{3r^2f^3}{2\tilde{\alpha}}\bigg(\frac{1}{\ell^2}+\frac{3}{4\tilde{\alpha}}\bigg)
 +\frac{D^4r^2}{3}\bigg(\frac{G}{\tilde{\alpha}}\bigg)^{\frac{4}{3}}f^\frac{7}{3}+\frac{GC}{\tilde{\alpha}r^2}
 =0.$$ This is a nonlinear second order ordinary differential equation and so has not an analytic closed solution. To solve it we must use
 numerical method or perturbation
 series expansion method.  The latter method is used to solve the metric potential equation
 (\ref{euler}) in the subsequent section.
\section{Metric solutions and Discretization} As a physical
boundary condition to solve the equation (\ref{euler}) we choose
\begin{equation}\lim_{r\to\infty}f(r)\to1
\end{equation} which by substituting the equation (\ref{euler}) reads to the following condition.
\begin{equation}\label{cond1}D^4\bigg(\frac{G}{\tilde{\alpha}}\bigg)^\frac{4}{3}=\frac{9}{2\tilde{\alpha}}
\bigg(\frac{1}{\ell^2}+\frac{3}{4\tilde{\alpha}}\bigg).
\end{equation} To solve the equation (\ref{euler}) by regarding the above condition it is useful we define
\begin{equation}\label{cond2}\frac{1}{d^2}=\frac{1}{\ell^2}+\frac{3}{4\tilde{\alpha}}
,~~~x=\frac{r}{d},~~~\epsilon=\frac{GC}{\tilde{\alpha}},~~~~\beta^4=\frac{d^2}{\tilde{\alpha}}.\end{equation}
By substituting (\ref{cond1}) and (\ref{cond2}) into the equation
(\ref{euler}) we obtain
\begin{equation}\label{eq}2f^4\ddot{f}+f^3\dot{f}^2-\frac{4f^4(f-1)}{x^2}+4\beta^4f^3(f-1)+\frac{3}{2}
\beta^4x^2(f^\frac{7}{3}-f^3)+
\frac{\epsilon}{x^2}=0\end{equation} where dot denotes to
derivative with respect to $x.$ Now we choose $\epsilon$ to be
order parameter in the following perturbation series function.
\begin{equation}\label{ser}f(x)=1+\epsilon g(x)+\epsilon^2 h(x)+\mathcal{O}(\epsilon^3).\end{equation}
 By substituting (\ref{ser}) into (\ref{eq})  and solving order by order of the metric equation (\ref{eq})
 we obtain $g(x)$ and $h(x)$.  First and second order of the equation (\ref{ser}) against
 $\epsilon$ order parameter are obtained respectively as follows.
 \begin{equation}\label{g}\ddot{g}(x)+\bigg(2\beta^4-\frac{\beta^4x^2}{2}-\frac{2}{x^2}\bigg)g(x)+\frac{1}{2x^2}=0
 \end{equation} and \begin{equation}\label{h}\ddot{h}(x)+\bigg(2\beta^4-\frac{\beta^4x^2}{2}-\frac{2}{x^2}\bigg)h(x)
 \end{equation}$$
 =-\frac{1}{2}\dot{g}^2(x)+\frac{2}{x^2}g(x)+
 \bigg(2\beta^4-\frac{11x^2\beta^4}{12}\bigg)g^2(x).$$ We do not bring here higher order terms of series form of the equation
 (\ref{eq}) because if we solve just the zero order equation (\ref{g}) then higher order solutions can be obtained easily step by
 step after where by substituting zero order solution $g(x)$ into the right hand side of the equation (\ref{h}) and obtain their particular
 solutions so on. Our strategy to solve (\ref{g}) is as follows:
 At first step we obtain asymptotically behavior of the function  $g(x)$ for  large distances $x>>1$ for which (\ref{g}) reads
 \begin{equation}\ddot{g}+(2\beta^4-\beta^4x^2/2)g(x)\approx0\end{equation}
 which has the following solutions in terms of the Hermite
 polynomials.
 \begin{equation}\label{hermit} g_n^{\infty}(y)\approx \frac{e^{-y^2/2}H_n(y)}{\Gamma_n}\end{equation}
by setting
\begin{equation}\label{beta}y=\beta
x/2^\frac{1}{4},~~~\beta^2=\frac{(2n+1)}{2\sqrt{2}}\end{equation}
where $\Gamma_n$ is normalization coefficient so that
\begin{equation}\Gamma_n=2^\frac{n}{2}\pi^\frac{1}{4}(n!)^\frac{1}{2}\end{equation} and $n=0,1,2,3,4,\cdots$ are order of the Hermite polynomials
$H_n(y)$.
At second step we substitute (\ref{hermit}), (\ref{beta})
 and definition \begin{equation}g(y)=g_n^{\infty}(y)F(y)\end{equation} into the equation (\ref{g}) to obtain differential
 equation for unknown $F(y)$ as follows.
 \begin{equation}\label{Fy}\frac{d^2F(y)}{dy^2}+2\bigg(-y+\frac{2nH_{n-1}(y)}{H_n(y)}\bigg)\frac{dF(y)}{dy}-\frac{2F(y)}{y^2}+
\frac{1}{2g^{\infty}_n(y)y^2}=0
 \end{equation} in which we used the identity $2nH_{n-1}(y)=\frac{dH_n(y)}{dy}$ for the Hermite polynomials
 and we supposed $F(y)$ gives behavior of the metric function $g(y)$ at $y\to0$ limits. It is easy to check that in
 $y\to0$ limits
 we have
  \begin{equation}\label{lim1}\lim_{y\to0} \frac{2(2n)H_{2n-1}(y)}{H_{2n}(y)}=0,~~~~\lim_{y\to0}g_{2n}^{\infty}(y)=\frac{H_{2n}(0)}{\Gamma_{2n}}=
  \frac{(-1)^n(2n)!}{n!\Gamma_{2n}}\equiv\frac{1}{e_{2n}}\end{equation}
  and
  \begin{equation}\label{lim2}\lim_{y\to0} \frac{2(2n+1)H_{2n}(y)}{H_{2n+1}(y)}=\frac{1}{y},~~~~\lim_{y\to0}g_{2n+1}^{\infty}(y)\approx
  \frac{2(2n+1)H_{2n}(0)}{\Gamma_{2n+1}}y\equiv\frac{y}{o_{2n+1}}
  \end{equation} respectively.
   By substituting these limits into the equation (\ref{Fy}) we obtain \begin{equation}\label{Fye}
   \frac{d^2F_e(y)}{dy^2}-\frac{2F_e(y)}{y^2}+\frac{e_{2n}}{2y^2}=0
 \end{equation}
 and
  \begin{equation}\label{Fyo}\frac{d^2F_o(y)}{dy^2}+\frac{2}{y}\frac{dF_o(y)}{dy}-\frac{2F_o(y)}{y^2}+
  \frac{o_{2n+1}}{2y^3}=0
 \end{equation}  in $y\to0$ limits respectively. One can obtain solutions of the equations (\ref{Fye}) and (\ref{Fyo}) respectively as follows.
 \begin{equation} F_e(y)=\frac{e_{2n}}{2}+\frac{\sigma_1}{y}+\sigma_2y^2\end{equation} and
 \begin{equation}F_o(y)=\frac{o_{2n+1}}{2y}+\frac{\eta_1}{y^2}+\eta_2y\end{equation} where $\sigma_{1,2}$
 and $\eta_{1,2}$
  are integral constants and they
 should be fixed by physical characteristics of local gravitational objects for instance mass and charge of black holes. Now we
 are in position to write
 explicit form of the metric potential (\ref{ser}) which up to second order terms become  \begin{equation}\label{fev}f_{2n}(y)=
 1+\frac{\epsilon e^{-y^2/2}}{\Gamma_{2n}}H_{2n}(y)\bigg(\frac{e_{2n}}{2}+\frac{\sigma_1}{y}+\sigma_2y^2\bigg)
 \end{equation}
 and
\begin{equation}\label{fodd}f_{2n+1}(y)=
 1+\frac{\epsilon e^{-y^2/2}}{\Gamma_{2n+1}}H_{2n+1}(y)\bigg(\frac{o_{2n+1}}{2y}+\frac{\eta_1}{y^2}+\eta_2y\bigg)
 \end{equation} respectively.
  We know from (\ref{lim1}) and (\ref{lim2}) \begin{equation}\label{her}H_{2n}(0)=(-1)^n\frac{(2n)!}{n!},~
 ~~\lim_{y\to0}H_{2n+1}(y)\sim\frac{2(-1)^n(2n+1)!}{n!}y.
 \end{equation} Now we apply to fix integral constants $\sigma_{1,2}$ and $\eta_{1,2}$ as follows.
 By substituting (\ref{cond2}), (\ref{beta}) and (\ref{her}), the metric potentials (\ref{fev}) and (\ref{fodd}) read
 \begin{equation}\label{even}\lim_{r<<d}
  f_{2n}(r)\sim1+\frac{\epsilon}{2}-\frac{2M}{r}+\frac{r^2}{\ell^2}\end{equation} and \begin{equation}
  \label{odd}\lim_{r<<d}
  f_{2n+1}(r)\sim1+\frac{\epsilon}{2}-\frac{2M}{r}+\frac{r^2}{\ell^2}\end{equation}
  respectively where we set
  \begin{equation}\label{sigma1}\sigma_1=-\frac{2^\frac{3}{4}\beta n!\Gamma_{2n}}{(-1)^n(2n)!\epsilon}\frac{M}{d}=
  -\frac{2^{n-1}n!}{(-1)^n}\sqrt{\frac{(2n+1)\sqrt{\pi}}{(2n)!}}\bigg(\frac{M\tilde{\alpha}}{CG}\bigg)\sqrt{\frac{4}{\ell^2}+\frac{3}{\tilde{\alpha}}}
 , \end{equation}\begin{equation}\label{sigma2}\sigma_2=\frac{\sqrt{2}n!\Gamma_{2n}}{\beta^2\epsilon
  (-1)^n(2n)!}\frac{d^2}{\ell^2}=\frac{16n!2^n}{(2n+1)(-1)^n}\sqrt{\frac{\sqrt{\pi}}{(2n)!}}\frac{\tilde{\alpha}^2}{CG(4\tilde{\alpha}+3\ell^2)}
  \end{equation} and \begin{equation}\label{eta1}\eta_1=-\frac{\beta n!\Gamma_{2n+1}}{\epsilon(-1)^n(2n+1)!2^\frac{1}{4}}\frac{M}{d}
  =-\frac{2^{n-1}n!}{(-1)^n}\sqrt{\frac{\sqrt{\pi}}{2(2n)!}}\bigg(\frac{M\tilde{\alpha}}{CG}\bigg)\sqrt{\frac{4}{\ell^2}+\frac{3}{\tilde{\alpha}
  }}\end{equation},\begin{equation}\label{eta2}\eta_2=\frac{n!\Gamma_{2n+1}}{\sqrt{2}\epsilon\beta^2(-1)^n(2n+1)!}
  \frac{d^2}{\ell^2}=\frac{n!2^{n+3}}{(2n+1)(-1)^n}\sqrt{\frac{2\sqrt{\pi}}{(2n+1)!}}\frac{\tilde{\alpha}^2}{CG(4\tilde{\alpha}+3\ell^2)}.\end{equation}
By comparing the above identities one can infer that
\begin{equation}\frac{\sigma_1}{\eta_1}=\frac{\sigma_2}{\eta_2}=\sqrt{2(2n+1)}.\end{equation}
It is easy to see that for fixed values of parameters of the model
called as $\{M,\ell,\tilde{\alpha}, C, G\}$ one can call
dimensionless eigenvalues for mass and AdS radiuses of the
discretized (quantum) 4DAdSGB black hole metric surrounded by NG
cloud of string as follows.
\begin{equation}\label{m}\sigma_1=-2m_{2n},~~~~\sigma_2=\frac{1}{\ell^2_{2n}},~~~\eta_1=-2m_{2n+1},~~~\eta_2=\frac{1}{\ell_{2n+1}^2}.\end{equation}
Looking at the relations (\ref{cond2}), (\ref{beta}) and metric
solutions (\ref{even}) and (\ref{odd}), one can infer that the
quantized distance
\begin{equation}d_n=\sqrt{\frac{\tilde{\alpha}}{2}}\bigg(n+\frac{1}{2}\bigg)\end{equation}
is a particular region for a fixed regularized GB parameter
$\tilde{\alpha}$ where the quantum behavior of this black hole
appears but not for distances less than it. In fact conditions
given by (\ref{sigma1}),  (\ref{sigma2}), (\ref{eta1}) and
(\ref{eta2}) are quantization conditions on the black hole
solution. They make the obtained black hole metric solution be
discretized. To end of this section we substitute (\ref{m}) into
the metric solutions (\ref{fev}) and (\ref{fodd}) to obtain exact
form of the discretized metric for the 4DAdSGB black hole
surrounded with string cloud. In this sense we obtain discretized
metric solutions for even and odd Hermite eigenvalues as follows.
\begin{equation}\label{f2n}f_{2n}(y)=1+\frac{n!}{(-1)^n(2n)!} e^{-y^2/2}H_{2n}(y)\bigg(1+\frac{\epsilon}{2}-
\frac{2m_{2n}}{y}+\frac{y^2}{\ell_{2n}^2}\bigg)
\end{equation} and
\begin{equation}\label{f2n1}f_{2n+1}(y)=1+\frac{n!}{(-1)^n(2n+1)!} e^{-y^2/2}H_{2n+1}(y)\bigg(1+\frac{\epsilon}{2}-
\frac{2m_{2n+1}}{y}+\frac{y^2}{\ell^2_{2n+1}}\bigg).\end{equation}
We now study thermodynamics behavior of the quantum black hole for
different quantum states $n=0,1,2,3,\cdots$ as follows.
\section{Thermodynamics of eigen states} Using the eigen (discretized) metric potentials
(\ref{f2n}) and (\ref{f2n1}) one can obtain eigen states for event
horizons by solving $f_{n}(y_+)=0$ for $even$ and $odd$ states as
follows.
\begin{equation}\label{mass1}m_{2n}=\frac{y_+}{2}\bigg(1+\frac{\epsilon}{2}+\frac{8\pi}{3}p_{2n}y_+^2+\frac{(-1)^n
(2n)!e^{y_+^2/2}}{n!H_{2n}(y_+)}\bigg)\end{equation} and
\begin{equation}\label{mass2}m_{2n+1}=\frac{y_+}{2}\bigg(1+\frac{\epsilon}{2}+\frac{8\pi}{3}p_{2n+1}y_+^2+\frac{(-1)^n
(2n+1)!e^{y_+^2/2}}{n!H_{2n+1}(y_+)}\bigg)\end{equation}
 where dimensionless ADM quantum mass of the black hole $m_{2n}$ and $m_{2n+1}$ are called as
 eigen enthalpy of the black hole in the extended thermodynamics of the black holes and they are described versus the black hole horizon $y_+$ for
 $even$ and $odd$ states respectively. In context of the extended phase of black hole thermodynamics the
eigen radiuses of the AdS  space treat as thermodynamic eigen
pressure as \begin{equation}p_{2n}=\frac{3}{8\pi
\ell^2_{2n}},~~~p_{2n+1}=\frac{3}{8\pi\ell^2_{2n+1}}\end{equation}
and corresponding  eigen thermodynamics volume is obtained from
the equation $v_n=\frac{\partial m_n}{\partial p_n}$ which by
substituting (\ref{mass1}) and (\ref{mass2}) reads
\begin{equation}v_{n}=\frac{4\pi}{3}y_{+}^3.\end{equation}
It is easy to see that the above quantized thermodynamics eigen
volume is equal to geometrical volume of the discretized black
hole and corresponding event horizon radius $y_+$ should be
substituted from $f_n(y_+)=0$ for $even$ and $odd$ eigen states
respectively. Applying the definition of the black hole Hawking
temperature $T=\frac{1}{4\pi}\frac{df(y)}{dy}\big|_{y=y_+}$ we
obtain eigen temperatures for $even$ and $odd$ states as follows.
\begin{equation}T_{2n}=\frac{1}{4\pi y_+}+\frac{y_+}{4\pi}-\frac{nH_{2n-1}(y_+)}{\pi H_{2n}(y_+)}+\frac{n! e^{-y^2_+/2}H_{2n}(y_+)}{4\pi(-1)^n
(2n)!}\bigg(\frac{1+\frac{\epsilon}{2}+8\pi
p_{2n}y_+^2}{y_+}\bigg)\end{equation} and
\begin{equation}T_{2n+1}=\frac{1}{4\pi
y_+}+\frac{y_+}{4\pi}\end{equation}$$-\frac{(2n+1)H_{2n}(y_+)}{2\pi
H_{2n+1}(y_+)}
+\frac{n!e^{-y_+^2/2}H_{2n+1}(y_+)}{4\pi(-1)^n(2n+1)!}\bigg(\frac{1+\frac{\epsilon}{2}+8\pi
p_{2n+1}y_+^2}{y_+}\bigg).$$ Entropy of this quantum black hole is
obtained for $even$ and $odd$ eigen states respectively as
follows. \begin{equation}
 s_{2n}=\int_0^{y_+}\frac{1}{T_{2n}}\bigg(\frac{\partial m_{2n}}{\partial y_+}\bigg)dy_+=\int\Sigma_{p_{2n}}^\epsilon(y_+)dy_+\end{equation}
 and
\begin{equation}
 s_{2n+1}=\int_0^{y_+}\frac{1}{T_{2n+1}}\bigg(\frac{\partial m_{2n+1}}{\partial y_+}\bigg)dy_+=\int\Sigma_{p_{2n+1}}^\epsilon(y_+)dy_+\end{equation}
 in which we defined
 \begin{equation}\Sigma_{p_{2n}}^{\epsilon}(y_+)=\frac{\frac{1}{2}+\frac{\epsilon}{4}+4\pi p_{2n}y_+^2+\frac{(-1)^n
 (2n)!e^{y_+^2/2}}{2n!}\bigg(\frac{(1+y_+^2)H_{2n}(y_+)
 -4nH_{2n-1}(y_+)}{H^2_{2n}(y_+)}\bigg)}{\frac{1}{4\pi y_+}+\frac{y_+}{4\pi}-\frac{nH_{2n-1}(y_+)}{\pi H_{2n}(y_+)}+\frac{n! e^{-y^2_+/2}H_{2n}
 (y_+)}{4\pi(-1)^n
(2n)!}\bigg(\frac{1+\frac{\epsilon}{2}+8\pi
p_{2n}y_+^2}{y_+}\bigg)}\end{equation} and
$$
 \Sigma^{\epsilon}_{p_{2n+1}}(y_+)$$\begin{equation}=\frac{\frac{1}{2}+\frac{\epsilon}{4}+4\pi p_{2n+1}y_+^2+
 \frac{(-1)^n(2n+1)!(1+y_+^2)e^{y^2_+/2}}{2n!H_{2n+1}(y_+)}
 -\frac{(-1)^n(2n+1)(2n+1)!y_+e^{y_+^2/2}H_{2n}(y_+)}{n!H^2_{2n+1}(y_+)}}{\frac{1}{4\pi y_+}+\frac{y_+}{4\pi}-\frac{(2n+1)H_{2n}(y_+)}{2\pi
 H_{2n+1}(y_+)}
 +\frac{n!e^{-y_+^2/2}H_{2n+1}(y_+)}{4\pi(-1)^n(2n+1)!}\bigg(\frac{1+\frac{\epsilon}{2}+8\pi
p_{2n+1}y_+^2}{y_+}\bigg)}.\end{equation} Applying the above
formulas one can obtain the Gibbs free energy of this quantum
black hole for $even$ and $odd$ states as
\begin{equation}G_n(y_+)=m_n(y_+)-T_n(y_+)s_n(y_+)\end{equation} which its form
is long and so does not shown here. As an example we continue our
thermodynamics study for ground state $n=0$ and first excited
state $n=1$ in the next subsection.
\subsection{Thermodynamics phase transition in ground state $n=0$}
By substituting $n=0$ into the obtained general form of equation
of states in the previous section we obtain enthalpy $m_0(y_+)$,
the Hawking temperature $T_0(y_+)$ and the entropy $s_0(y_+)$
respectively as follows.
\begin{equation}\label{m0}m_0(y_+)=\frac{y_+}{2}\bigg(1+\frac{\epsilon}{2}+\frac{8\pi y_+^2}{3}p_0+e^{y^2_+/2}
\bigg)\end{equation}
\begin{equation}\label{T0}T_0(y_+)=\frac{1}{4\pi y_+}+\frac{y_+}{4\pi}+\frac{e^{-y_+^2/2}}{4\pi}\bigg(\frac{1+
\frac{\epsilon}{2}+8\pi p_0y_+^2}{y_+}\bigg)\end{equation}
\begin{equation}s_0(y_+)=2\pi(e^{y_+^2/2}-1).\end{equation}
Looking at the temperature equation  (\ref{T0}) and comparing with
the ideal gas equation of state one can infer that the
corresponding specific volume in ground state become
\begin{equation}\nu_0=2y_+e^{-y^2_+/2}.\end{equation}
By eliminating $y_+$ between this specific volume equation and the
Hawking temperature (\ref{T0}) we can obtain PVT imperfect
equation of state of the quantum black hole in ground state as
follows.
\begin{equation}\label{p0}p_0(\nu_0)=-\frac{(1+\frac{\epsilon}{2})e^{L(\nu_0)}}{4\pi\nu_0}
+e^{\frac{\nu_0^2e^{-2L(\nu_0)}}{8}}\bigg(\frac{T_0e^{L(\nu_0)}}{\nu_0}-\frac{e^{2L(\nu_0)}}{2\pi\nu_0^2}-\frac{1}{8\pi}
\bigg)\end{equation} where we defined
\begin{equation}L(\nu_0)=\frac{1}{2}LambertW\bigg(-\frac{\nu_0^2}{2}\bigg)\end{equation}
  One can obtain critical point by
solving the equations $\frac{\partial p_0}{\partial\nu_0}=0$ and
$\frac{\partial^2 p_0}{\partial\nu_0^2}=0$ at constant
temperature. In fact these equations are equivalent with the
equations $\frac{\partial p_0}{\partial y_+}=0$ and
$\frac{\partial^2 p_0}{\partial y_+^2}=0$ at constant temperature
which lead to the following parametric critical points if we
substitute the pressure  $p_0(y_+)$ given by (\ref{T0}) into them.
\begin{equation}\label{crit}T_{0c}=\frac{y_{+c}^3(5+y_{+c}^2)}{4\pi(y^4_{+c}+2y_{+c}^2-1)},~~~p_{0c}=\frac{(3+y^4_{+c})e^{y^2_{+c}/2}}{
16\pi(y^4_{+c}+2y^2_{+c}-1)},~
~~\nu_{0c}=2y_{+c}e^{-y^2_{+c}/2}
\end{equation}
in which the critical radius of the event horizon in ground state
$y_{+c}$ is determined versus the string tension $\epsilon$ by
solving the following
equation\begin{equation}\label{epsilon}\epsilon=\frac{2-4y_{+c}^2-2y^4_{+c}+e^{y^2_{+c}/2}(2-5y^2_{+c}+4y^4_{+c}-y^6_{+c})}{y^4_{+c}+2y^2_{+c}-1}.
\end{equation}
We plotted diagrams of the above equation $\epsilon(y_{+c})$ and
all possible numerical values for critical points given by
(\ref{crit}) in figure 1. To continue study of thermodynamics of
this black hole we should choose sample numerical value from
diagrams of the figure 1 for instance $\epsilon=-1.$ In this case
we obtain
$$y_{+c}=0.7394220670,~~~\nu_{0c}=1.125118423,~$$\begin{equation}p_{0c}=0.2198245577,~~~T_{0c}=0.4547307702\end{equation}
 in which critical enthalpy of this black hole is obtained as
\begin{equation}m_{0c}=1.043056198
\end{equation}
which is useful to study Joule-Thomson expansion by plotting
isenthalpic $T-P$  diagrams. $p-v$ diagram in figure 2 shows a
small to large black hole phase transition which is happened at
temperature higher than the critical temperature $T_0>T_{0c}$ at
ground state $n=0.$ Other diagrams namely $p-y_+$ or $T-v$ predict
similar situations. We know that a thermodynamic system is stable
when its Gibbs free energy takes negative values. For this black
hole it is seen at figure 3 for $G-v$ or $G-y_+$ diagrams. In
these diagrams minimum point of the Gibbs free energy with
negative value is at bigger size black holes. This means that the
black hole participates at small to large black hole phase
transition. This is happened at pressures higher than the critical
pressure $p_0>p_{0c}.$ One can see from $G-v$ or $G-y_+$ diagrams
that the enormous black holes have not negative Gibbs energy which
means that this quantum black holes at ground state can not to
participate in the Hawking-Page phase transition where a black
hole evaporates and its final state reaches to a vacuum AdS space.
We will see this is possible in the first excited states by
plotting $G-v$ or $G-y_+$ curves of this black hole (see figure 7
in the next section).
\subsection{Thermodynamics phase transition in first excited state $n=1$}
By substituting $n=1$ into the obtained general form of equation
of states in the previous section we obtain enthalpy $m_1(y_+)$,
the Hawking temperature $T_1(y_+)$ and the entropy $s_1(y_+)$
respectively as follows.
\begin{equation}\label{m1}m_1(y_+)=\frac{y_+}{2}\bigg(1+\frac{\epsilon}{2}+\frac{8\pi y_+^2}{3}
p_1+\frac{e^{y^2_+/2}}{2y}\bigg)\end{equation}
\begin{equation}\label{T1}T_1(y_+)=\frac{y_+}{4\pi}+\frac{e^{-y_+^2/2}}{2\pi}\bigg(1+
 \frac{\epsilon}{2}+8\pi
p_1y_+^2\bigg)\end{equation}
\begin{equation}s_1(y_+)=-i\pi\sqrt{\frac{\pi}{2}}erf\bigg(\frac{iy_+}{\sqrt{2}}\bigg).\end{equation}
By looking at the temperature equation  (\ref{T1}) and by
comparing with the ideal gas equation of state one can infer that
the corresponding specific volume in first excited state become
\begin{equation}\label{nu1}\nu_1=4y_+^2e^{-y^2_+/2},~~~~y^{(\pm)}_{+}=\pm\sqrt{2L(\nu_1)}\end{equation}
where we defined
\begin{equation}L(\nu_1)=-LambertW\bigg(-\frac{\nu_1}{96}\bigg).\end{equation} By eliminating $y_+$ between this specific volume equation
(\ref{nu1}) and the Hawking temperature (\ref{T1}) we can obtain
PVT equation of state of the quantum black hole in first excited
state as follows.
\begin{equation}p^{(\pm)}_1(\nu_1)=\frac{192\pi L(\nu_1)T_1\mp48\sqrt{2}L(\nu1)^\frac{3}{2}-(1+
\frac{\epsilon}{2})\nu_1}{16\pi\nu_1 L(\nu_1)}.\end{equation}
 One can obtain critical point by
solving the equations $\frac{\partial p_1}{\partial\nu_1}=0$ and
$\frac{\partial^2 p_1}{\partial\nu_1^2}=0.$ We checked these
equations to obtain some real positive values for the critical
specific volume $\nu_{1c}$ and saw that there are complex
(analytic continuation) values for the critical event horizons
$y_{+c}$ which still give out real values for the specific volume
$\nu_{1c}.$ To exclude these analytic continuation horizon
radiuses we solve the equivalent critical points equations
$\frac{\partial p_1}{\partial y_+}=0$ and $\frac{\partial^2
p_1}{\partial y_+^2}=0$ to obtain  real positive values for the
critical radius of the horizons and the corresponding specific
volume of the black hole horizons as follows. After solving the
equations $\frac{\partial p_1}{\partial y_+}=0$ and
$\frac{\partial^2 p_1}{\partial y_+^2}=0$ and some simple
mathematical derivations we obtain general form of the critical
points against $y_{+c}$ as follows.
\begin{equation}\label{T1c}T_{1c}=\frac{y_{+c}^4+2y^2_{+c}-1}{4\pi y_{+c}^3},~~~p_{1c}=\frac{
(y^2_{+c}-1)e^{-y^2_{+c}/2}}{32\pi y^3_{+c}},~~~\nu_{1c}=4y^2_{+c}e^{-y^2_{+c}/2}
\end{equation}
where the critical radius of the event horizon $y_{+c}$ is
obtained from largest positive real root of the following
equation.
\begin{equation}\epsilon=-2-\frac{e^{y^2_{+c}/2}}{2}\bigg(y_{+c}-\frac{5}{y_{+c}}+\frac{2}{y^3_{+c}}\bigg).\end{equation}
This root will be correspond to exterior horizon of the black hole
which is visible for observers located at outside of the black
hole. We show all possible real roots of the above equation for a
given string tension $\epsilon$ in figure 4.  We continue our
study as numerically by choosing a particular value from the
figure 4 as follows.
\begin{equation}\label{-}
\nu_{1c}=2.198851498,~~~p_{1c}=0.0005220136425,~~~T_{1c}=0.2279369128,\end{equation}
for
\begin{equation}\epsilon=-1,~~~y_{+c}=1.984472401\end{equation}
where critical black hole enthalpy at $n=1$ state is obtained as
follows
\begin{equation}m_{1c}=2.304201063\end{equation}which
is applicable to study isenthalpic $T-p$ curves and Joule-Thomson
expansion of the system.
 All possible critical points given by the equations (\ref{T1c}) are
plotted versus the string tension $\epsilon$ at first excited
state $n=1$ in figure 4.
 By looking at the isobaric $T-v$ and $T-y_+$ curves in figure 5 one
 can understand that the temperature has local maximum (minimum) value for smaller (larger) black holes at
 pressures higher than the critical pressure.
  This means that the black hole participates at the small to large scale
 black hole phase transition for pressures higher than the critical
 pressure. Isobaric $T-v$ curves show that the black hole have
 maximum temperature at largest scale at $p>p_c.$ This diagram shows that the black hole can not to participate at the
 Hawking-Page phase transition apparently where the black hole evaporates and finally reaches to the Ads vacuum space.
 But by looking at $G-v$ or $G-y_+$ isobaric curves one can infer that negativity sign of
 the Gibbs free energy is happened at
 large black holes and huge sized black holes (see figures 7 and 8) in which the first small regions with
 $G<0$ is related to the small to large phase transition
 while the larger regions with $G<0$ is corresponds to the Hawking-Page phase transition \cite{Hawking}.
 One can obtain similar
 statements by looking at the isothermal $p-v$  curves given by  figure 6.
 In context of black holes thermodynamics the small to large scale black hole phase transition is confirmed that is similar
  to solid-fluid-gas phase transition in imperfect Van der Waals fluid.
In fact the diagrams show different behavior for the black hole in
the below and the above of the critical points of the phase space.
If we want to explain accurately, in fact the small to large black
hole phase transition is happened by connecting an unstable medium
size black hole near the maximum point of $p-v$ diagram in the
ground state (see figure 2) and the first excited state (see
figure 7). In any thermodynamic system always exist a phase with
minimum value for its Gibbs free energy. When two branches of
Gibbs free  energy at minimum state cross each other then the
phase transition is happened, while  if these two branches have
the same value of Gibbs free energy then this two phases are
called coexist. This swallowtail behavior indicates coexistence of
two different phases  which is happened in ground state $n=0$ (see
G-T curve at constant pressure $p_0>p_{0c}$ in figure 3) and at
first excited state $n=1$ (see G-T curve at constant pressure
$p_1>p_{1c}$ in figure 8). At end of this section we can
understand by looking the G-y or G-v curves given in the figures 3
and 8 where an unstable small black hole with $G>0$ reaches to an
stable large black hole with $G<0$. While an unstable huge black
hole with $G>0$ evaporate completely and reduces to a vacuum AdS
space with $G<0$ by participating in the Hawking-Page phase
transition.
\section{Conclusion}
According to the 4DAdS EGB gravity model \cite{ali} we obtained
discretized metric potentials for spherically symmetric static
quantum black holes surrounded with cloud of string which is
described by the Hermite polynomials (spherical quantum harmonic
oscillator). Thus we called our obtained solutions as quantized
(quantumized) black hole. We should point that our used
`quantization` adjective for this black hole metric solution is
different with the well known canonical quantization. In fact we
used this adjective because of two reasons as follows: (a) The
metric solution is obtained by the Hermite polynomials with
`discrete` eigenvalues. (b) The used EGB gravity as an higher
order derivative of metric theory originates from renormalization
of quantum fields propagating in curved space times. We used the
extended thermodynamic formalism to show that $4D$ AdS EGB black
holes surrounded by strings fluid mimics Van der Waals fluid
behavior. In extended thermodynamic paradigm, pressure of AdS
background spacetime which is affect on a local black hole metric
is described by negative cosmological constant and so its
conjugate variable plays the role of thermodynamic volume of the
black hole. To examine the analogy between $4D$ AdS EGB black
holes surrounded by cloud of strings with ordinary liquid-gas
thermodynamic system we derived analytical solutions of the
critical points in different eigenstates which are fixed by the
string tension. With in detail we studied thermodynamics  of this
black hole at the ground state and first excited state
respectively and obtained dependence of the black hole
thermodynamic phase transition to the black hole quantum states.
In this sense the black hole in ground state participates just at
the phase transition of small to large scale black holes if it has
small Gibbs energy while it participates at the Hawking-page phase
transition if it has large Gibbs free energy. On the other side
for a black hole with small Gibbs free energy when is at the first
excited state there is possibility of participation of the black
hole in both of the small to large and also the Hawking-Page phase
transitions. Also we checked diagrams of the isenthalpic P-T
curves and obtained that this black hole does not participates in
the Joule-Thomson expansion phenomena in ground state and first
excited state. As a future work we like to study canonical
quantization of this black hole by solving its Wheeler De Witt
wave equation and obtain its relation with the thermodynamic phase
transition by comparing with our obtained results in this work.

\begin{figure}[tbp]
 \centering {
 \includegraphics[width=6cm]{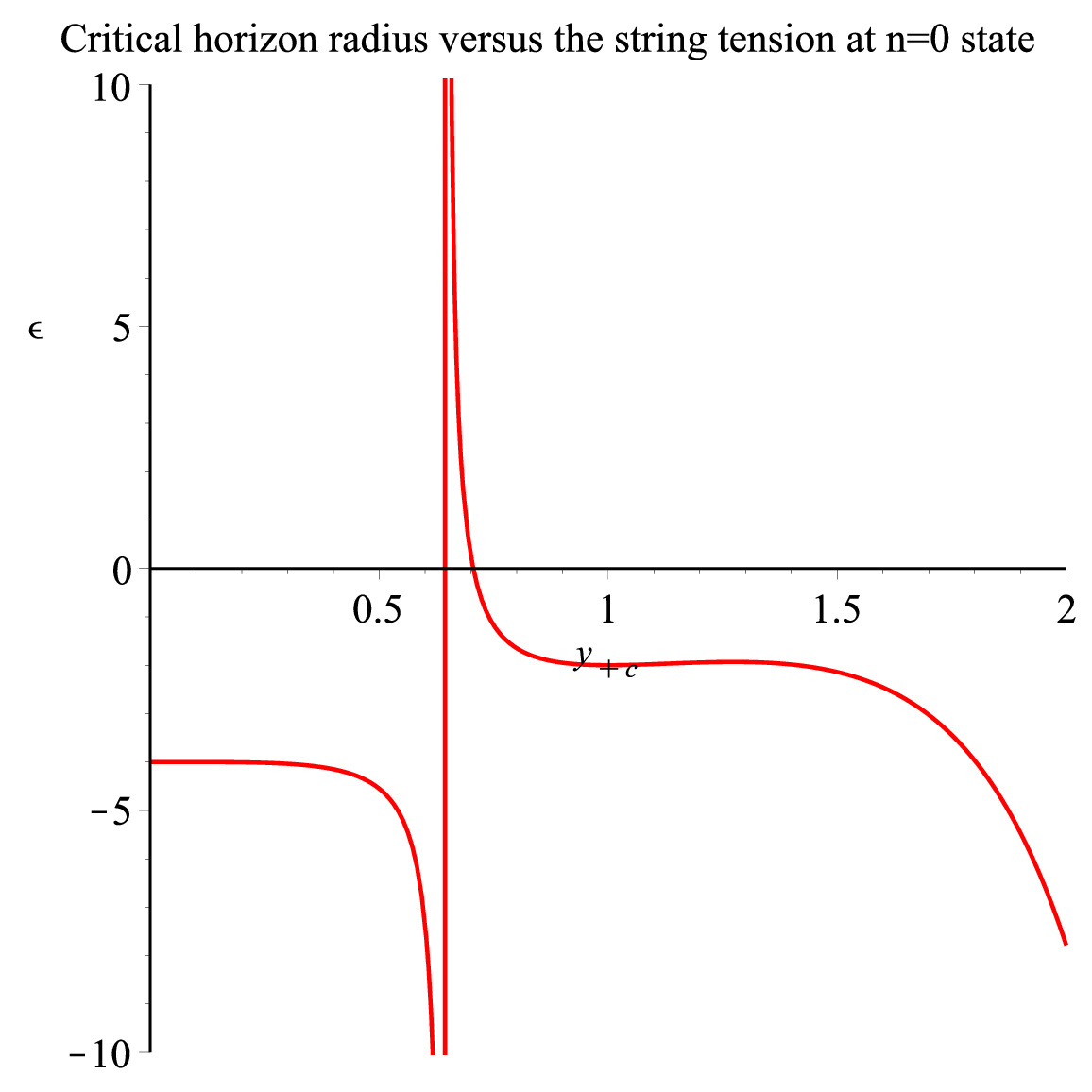}
 \includegraphics[width=6cm]{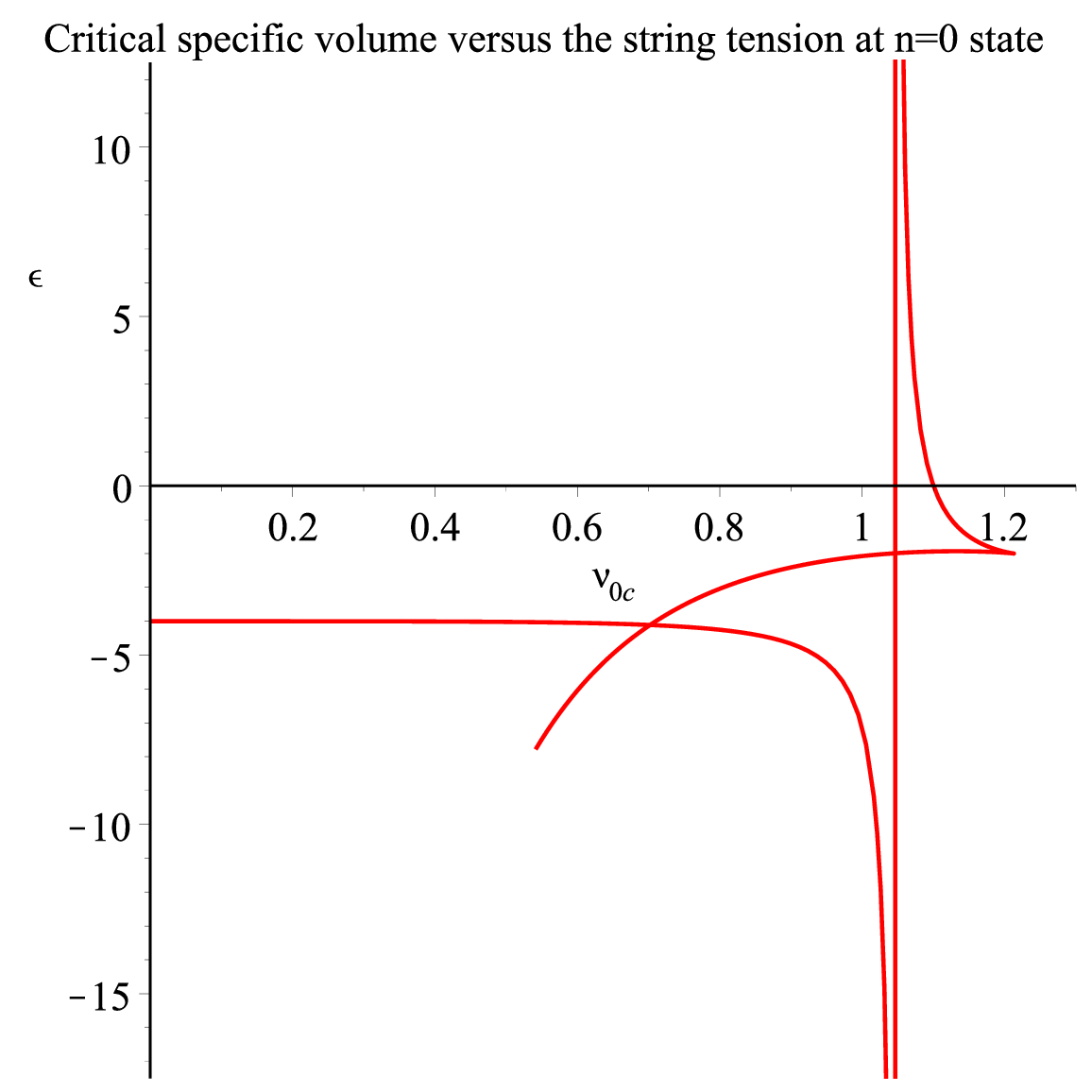}
 \includegraphics[width=6cm]{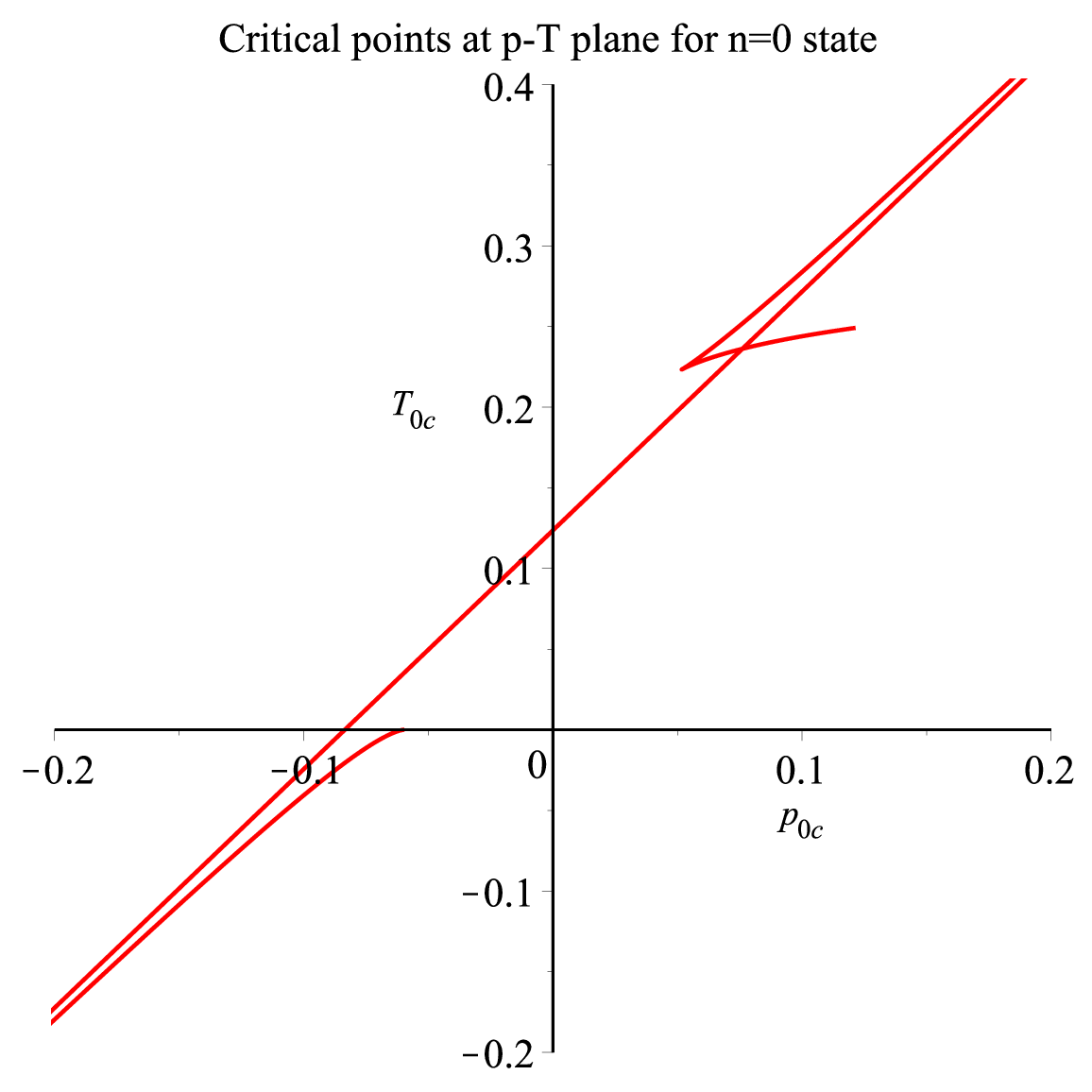}
 \includegraphics[width=6cm]{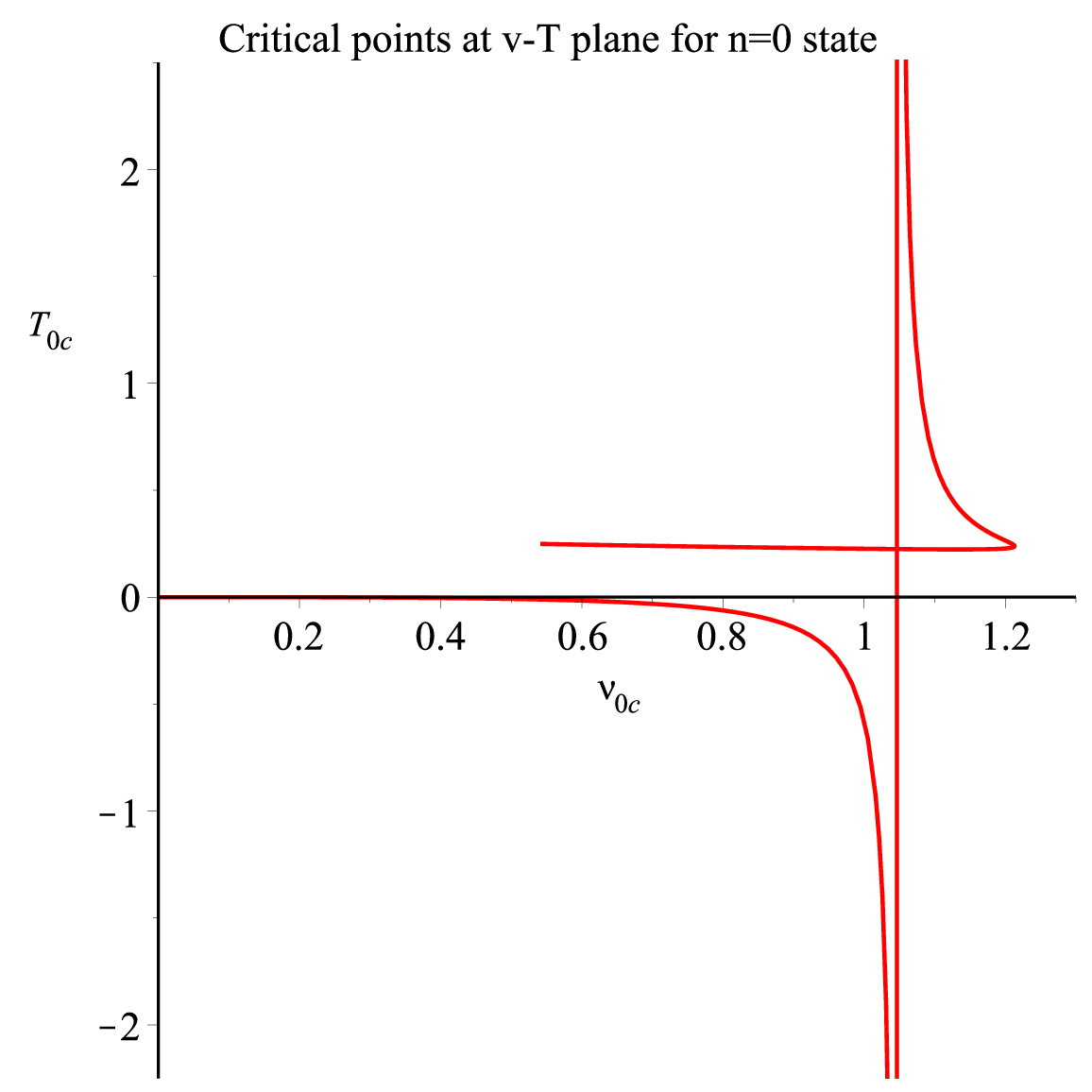}
 \includegraphics[width=6.cm]{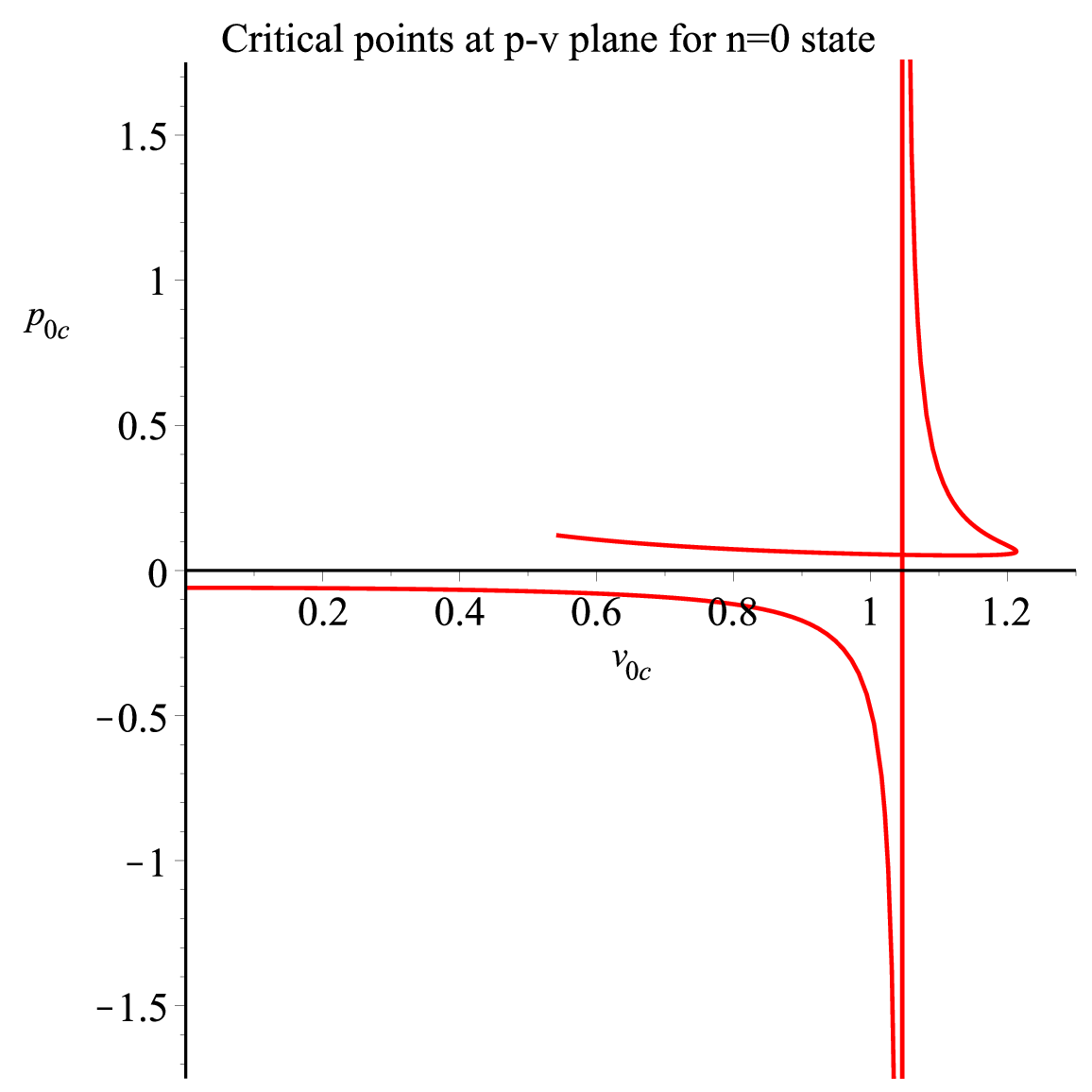}
 \includegraphics[width=6.cm]{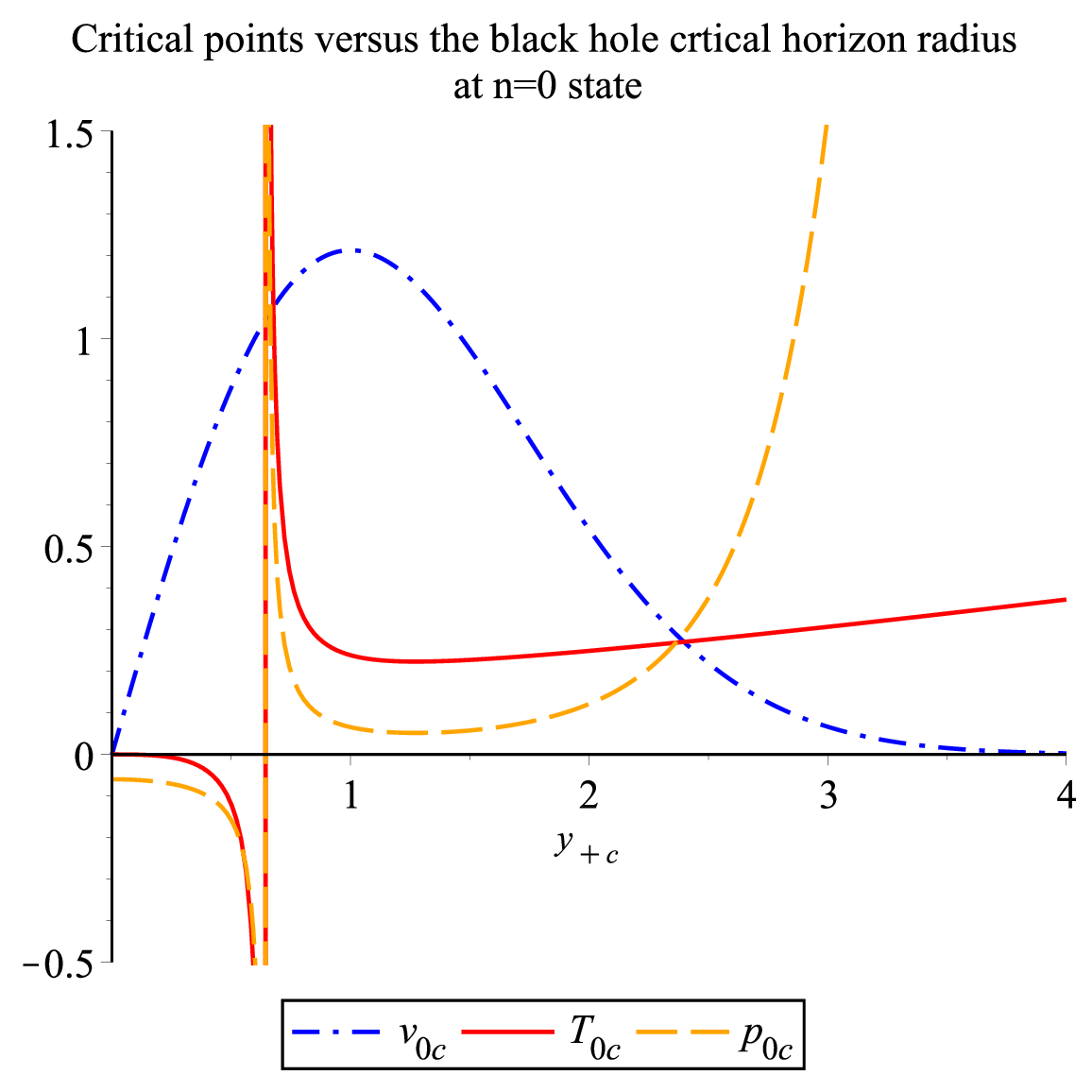}
}
 \caption{Numerical values of all possible critical points versus the string tension $\epsilon$ at ground state of the AdS 4DGB quantum black hole $n=0$}
\end{figure}
\begin{figure}[tbp]
\centering{\includegraphics[width=6cm]{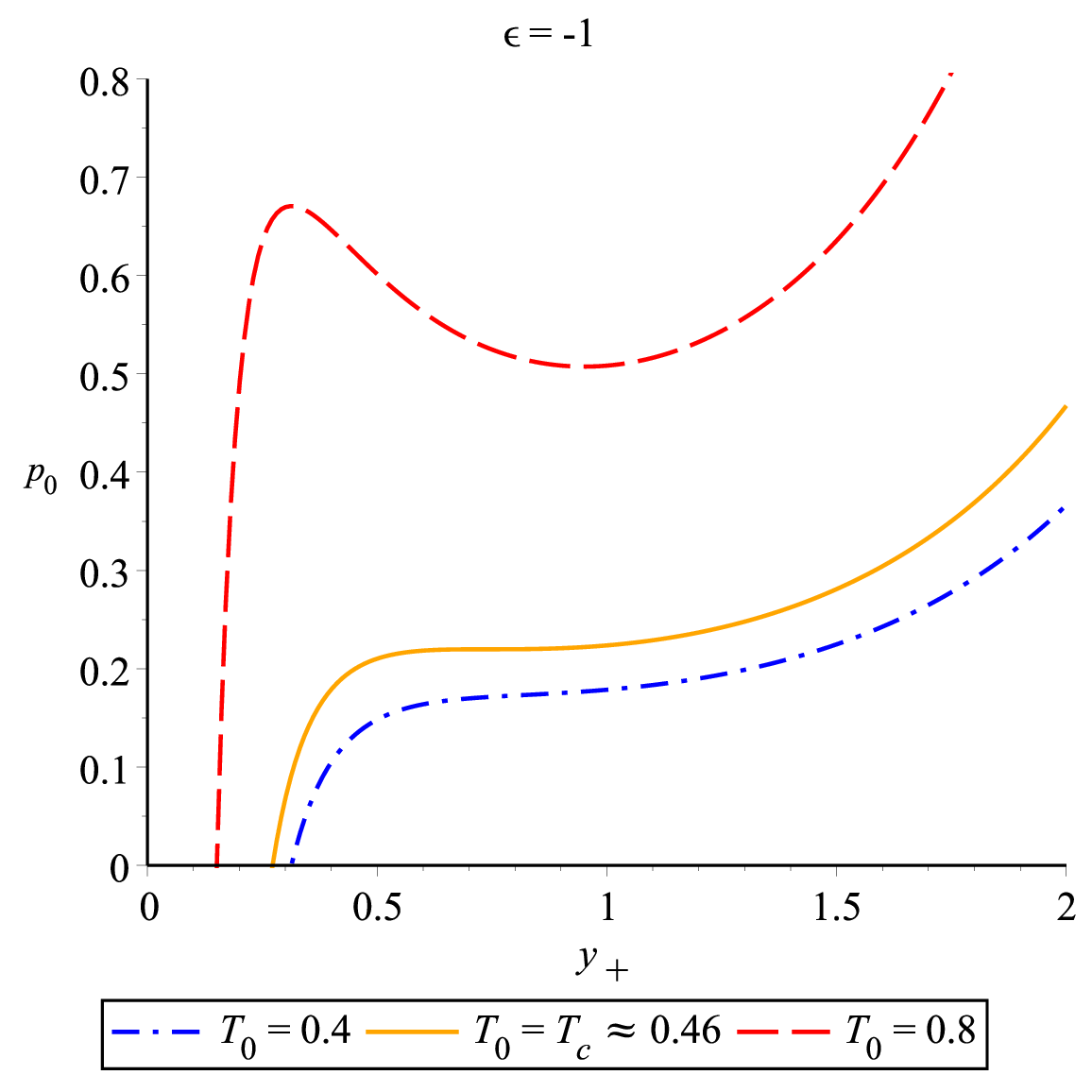}
\includegraphics[width=6cm]{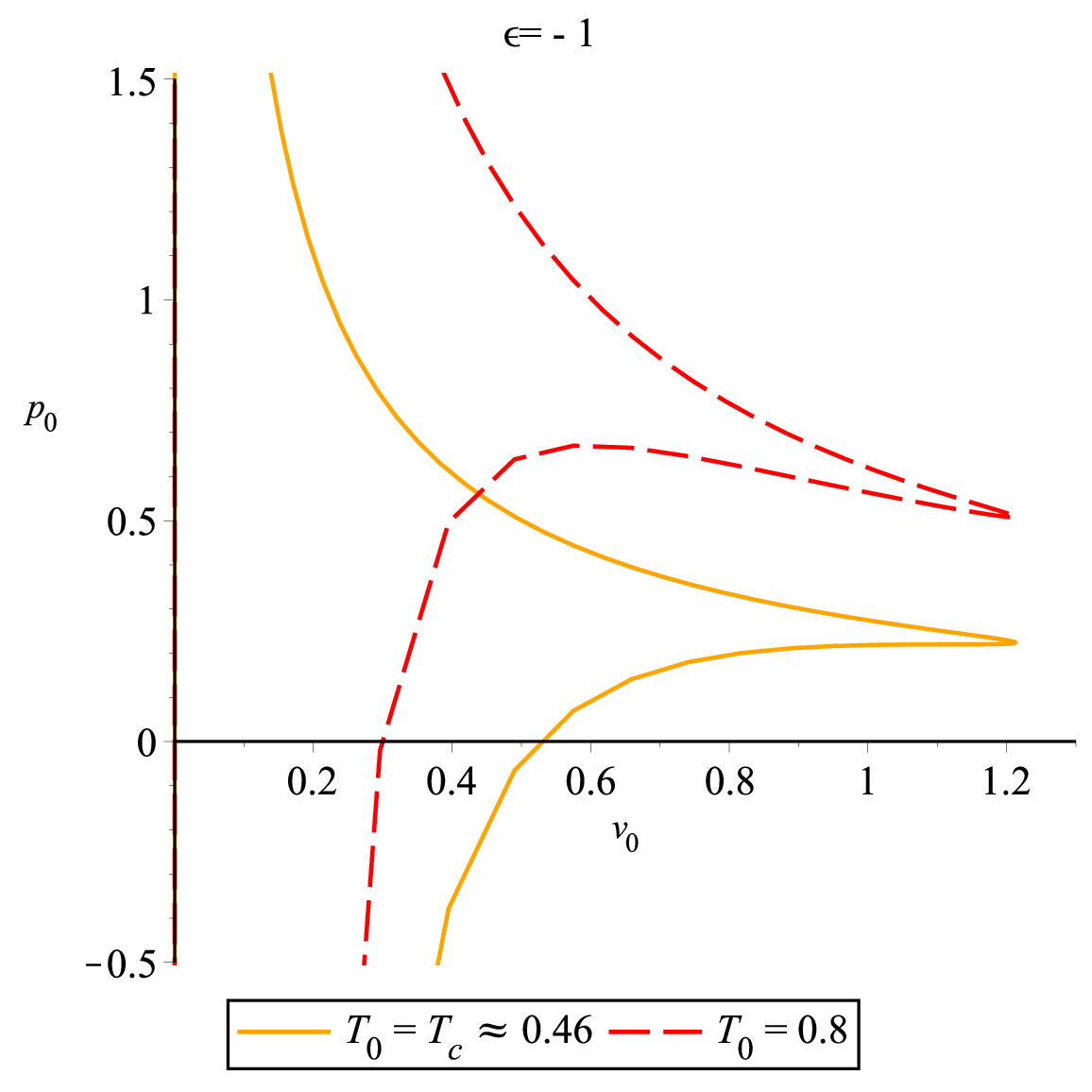}
\includegraphics[width=6cm]{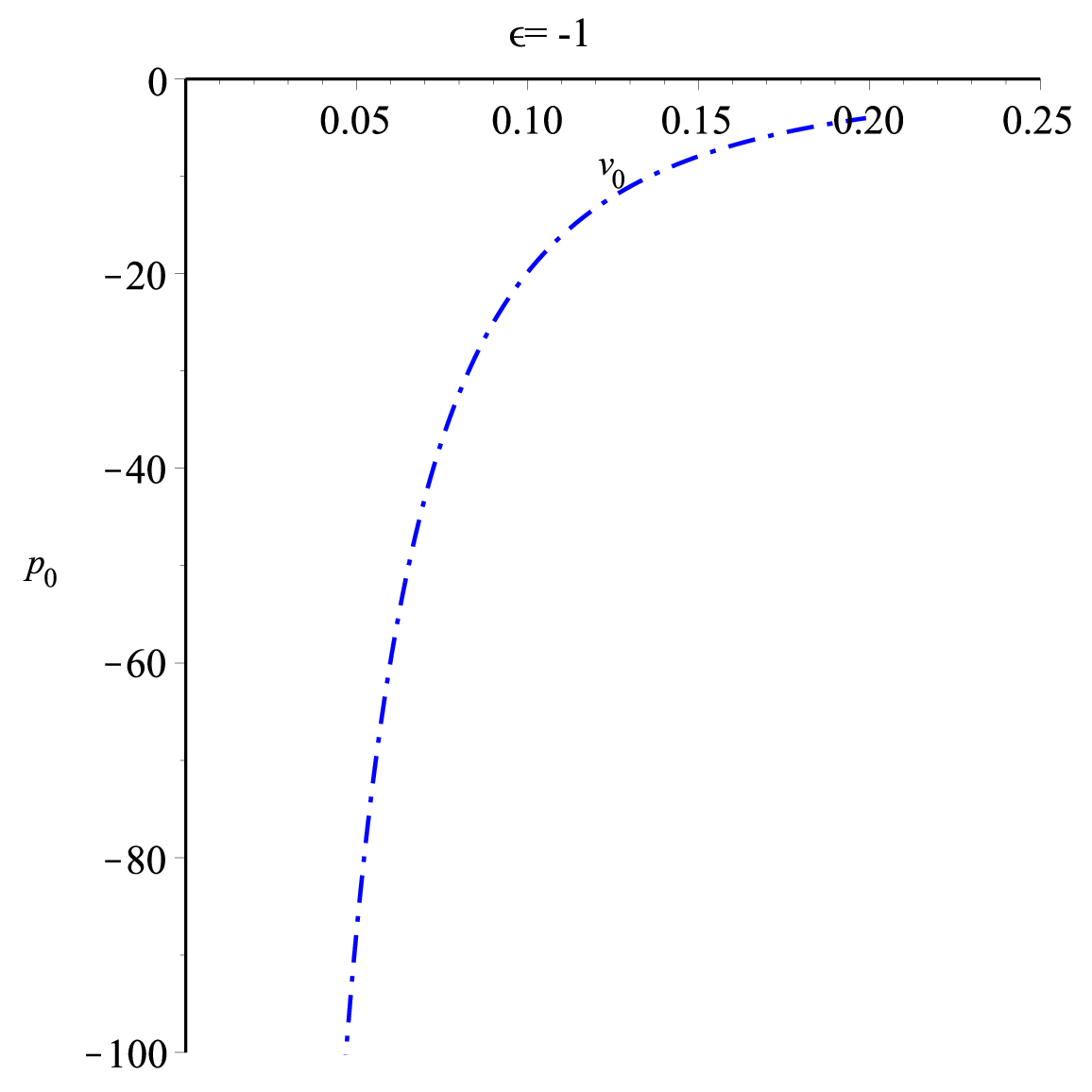}
\includegraphics[width=6cm]{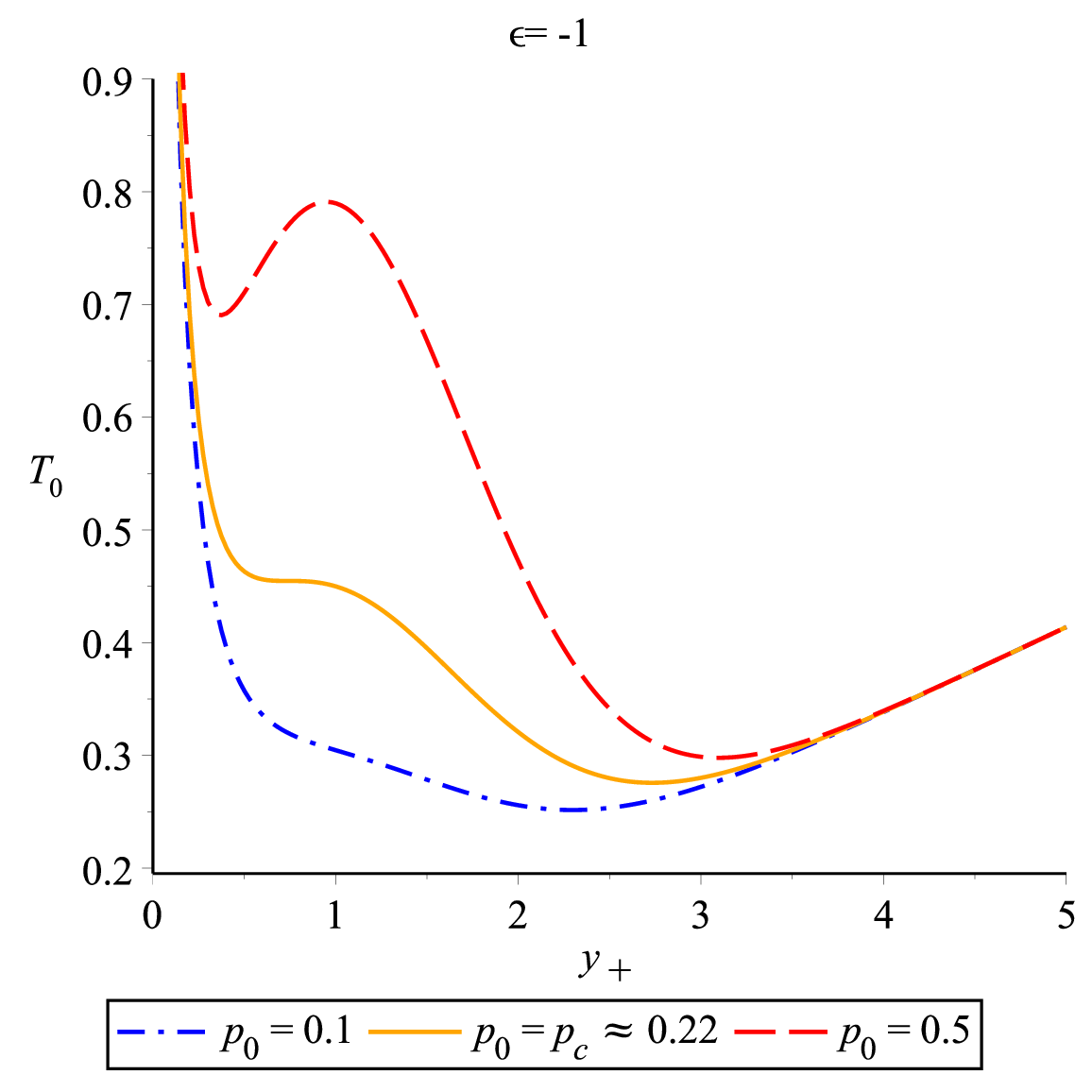}
\includegraphics[width=6cm]{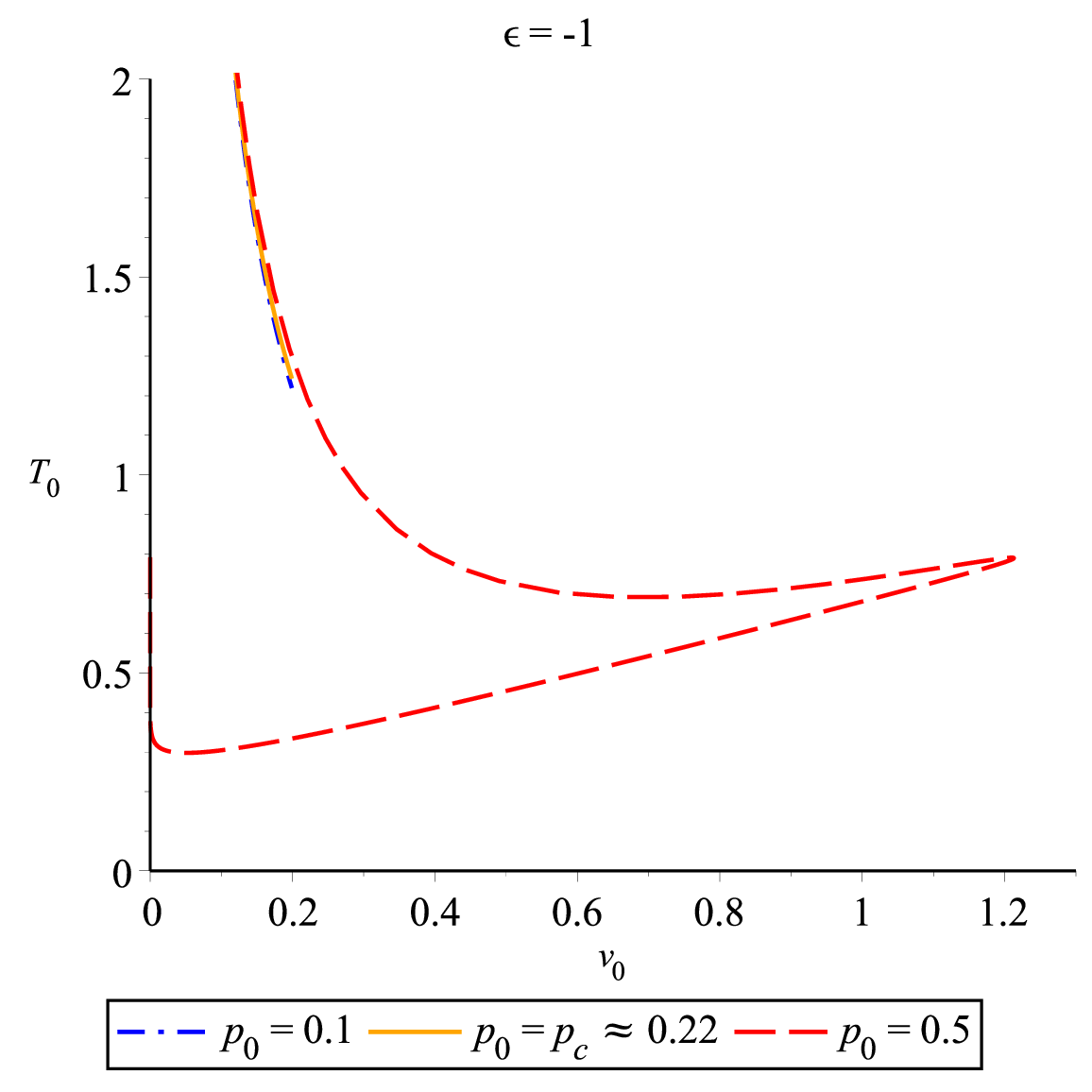}
}
 \caption{Diagrams for $p(v),p(y_+)$ and $T(v)$ at ground state n=0}
\end{figure}
\begin{figure}[tbp]
\centering{
\includegraphics[width=6.7cm]{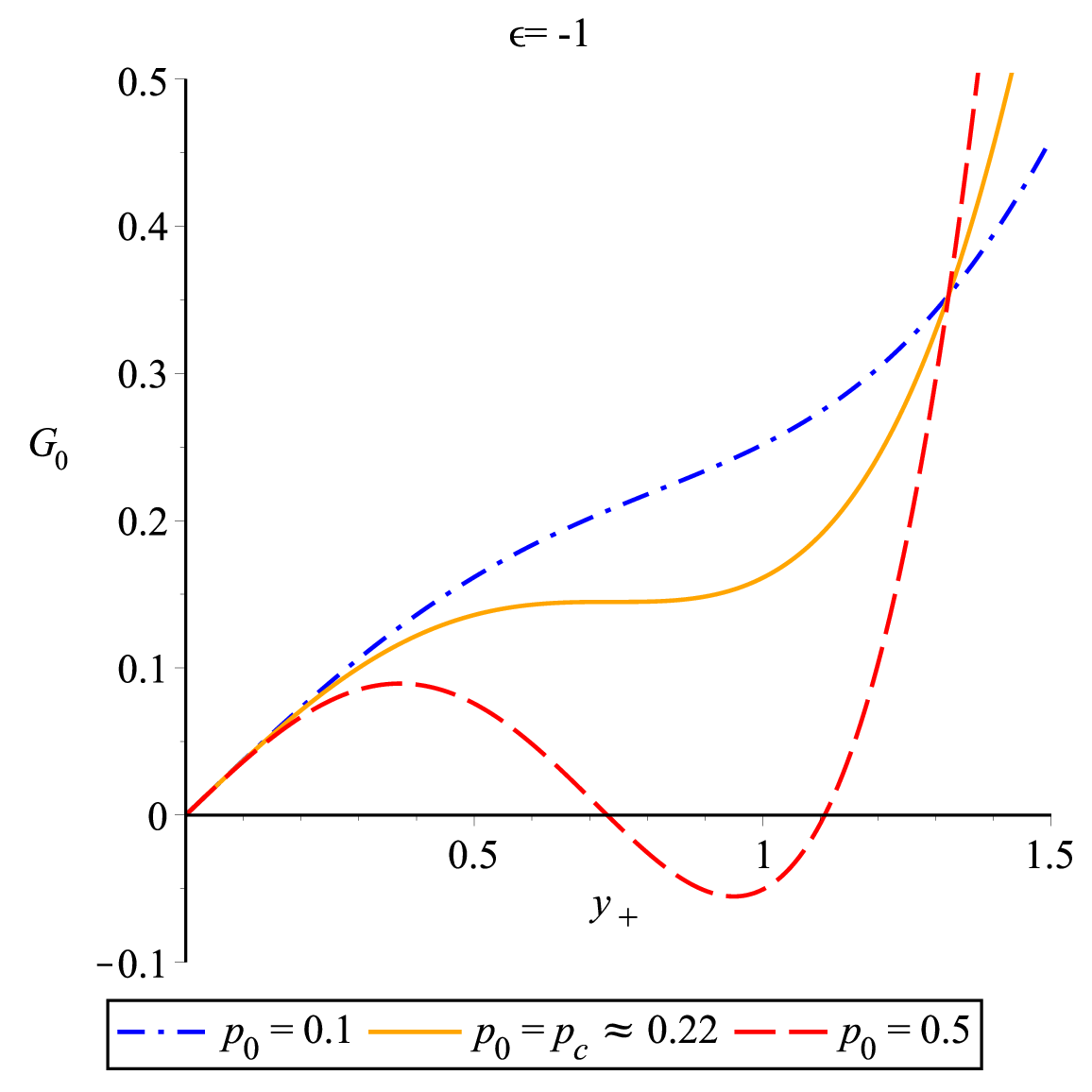}
\includegraphics[width=6.7cm]{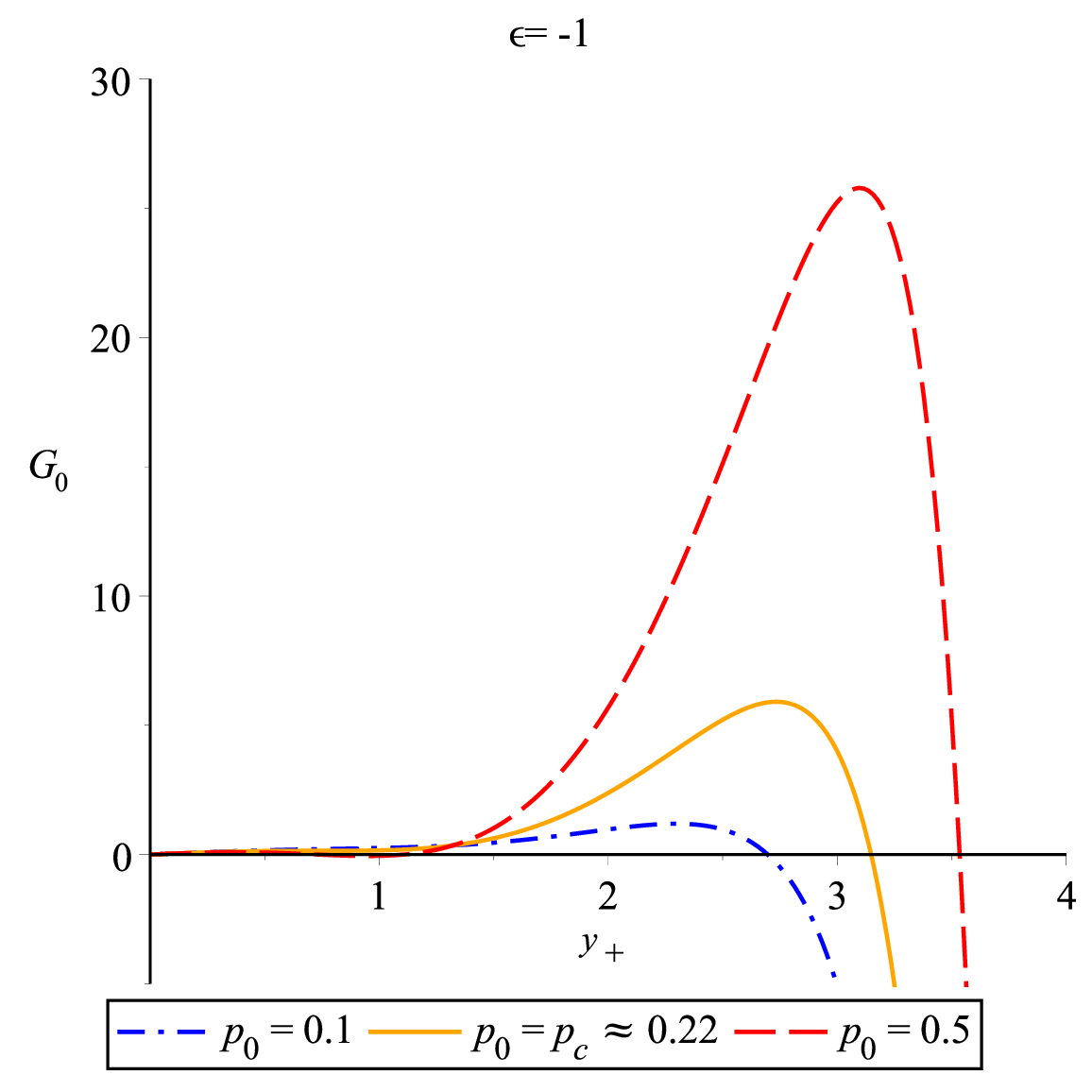}
\includegraphics[width=6.7cm]{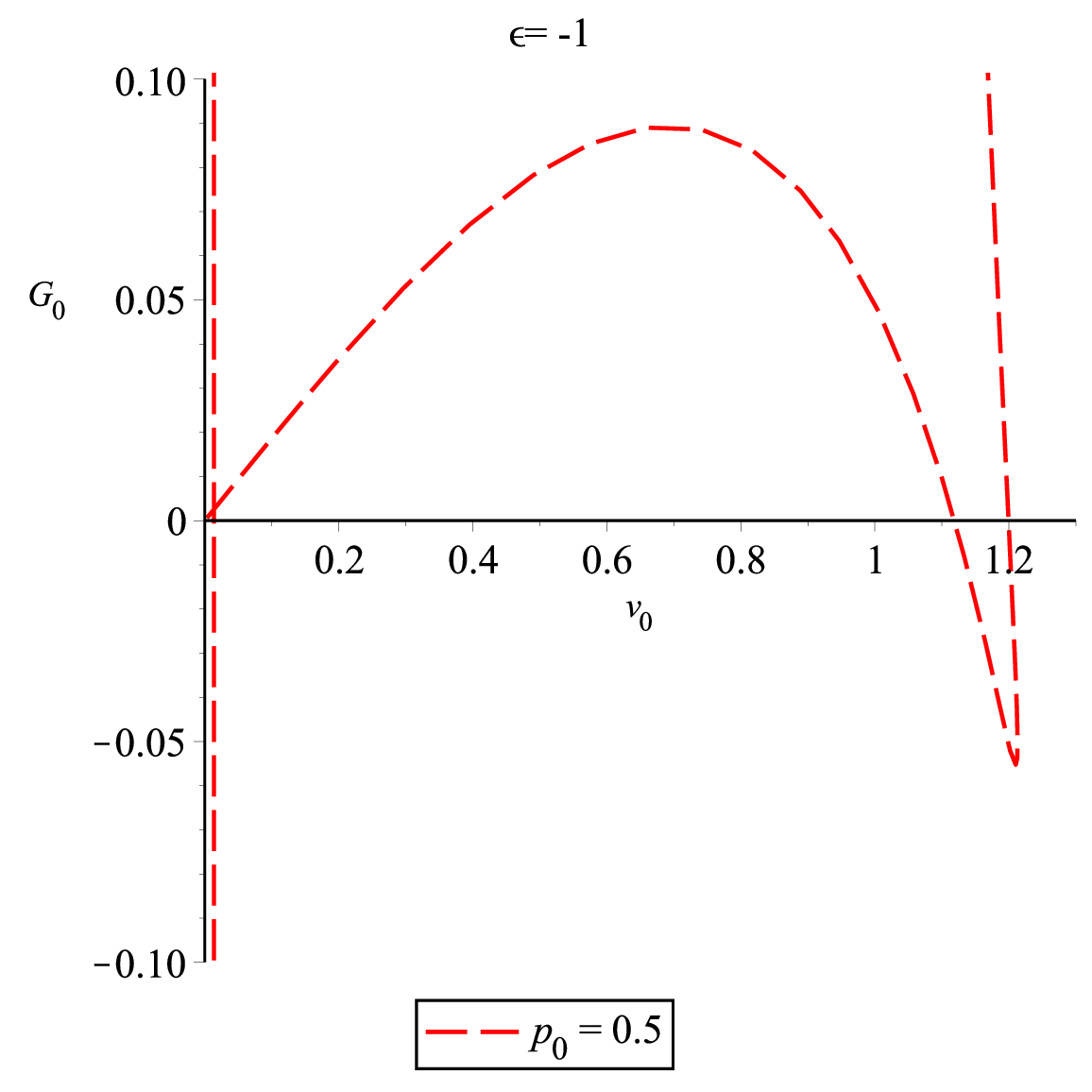}
\includegraphics[width=6.7cm]{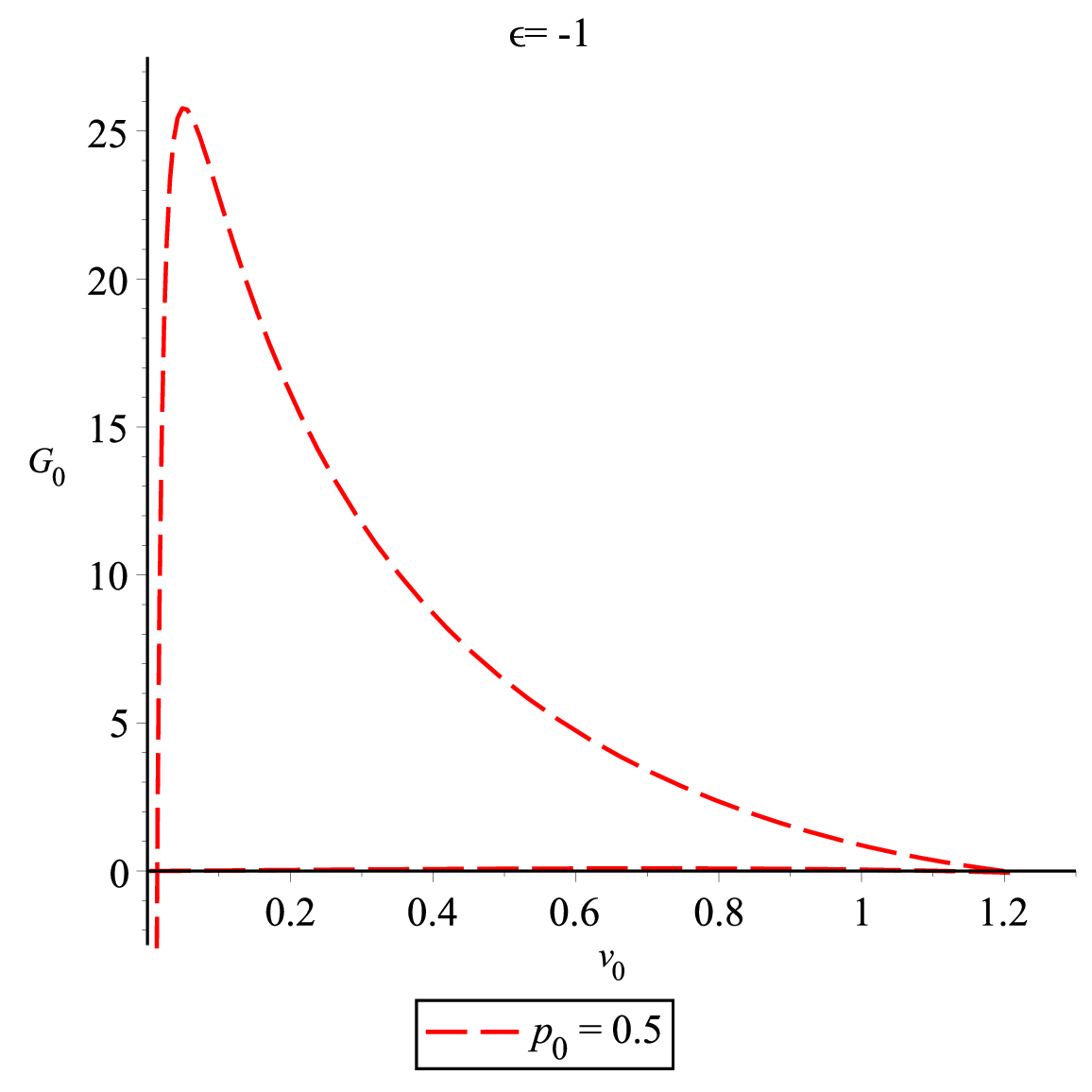}
\includegraphics[width=6.7cm]{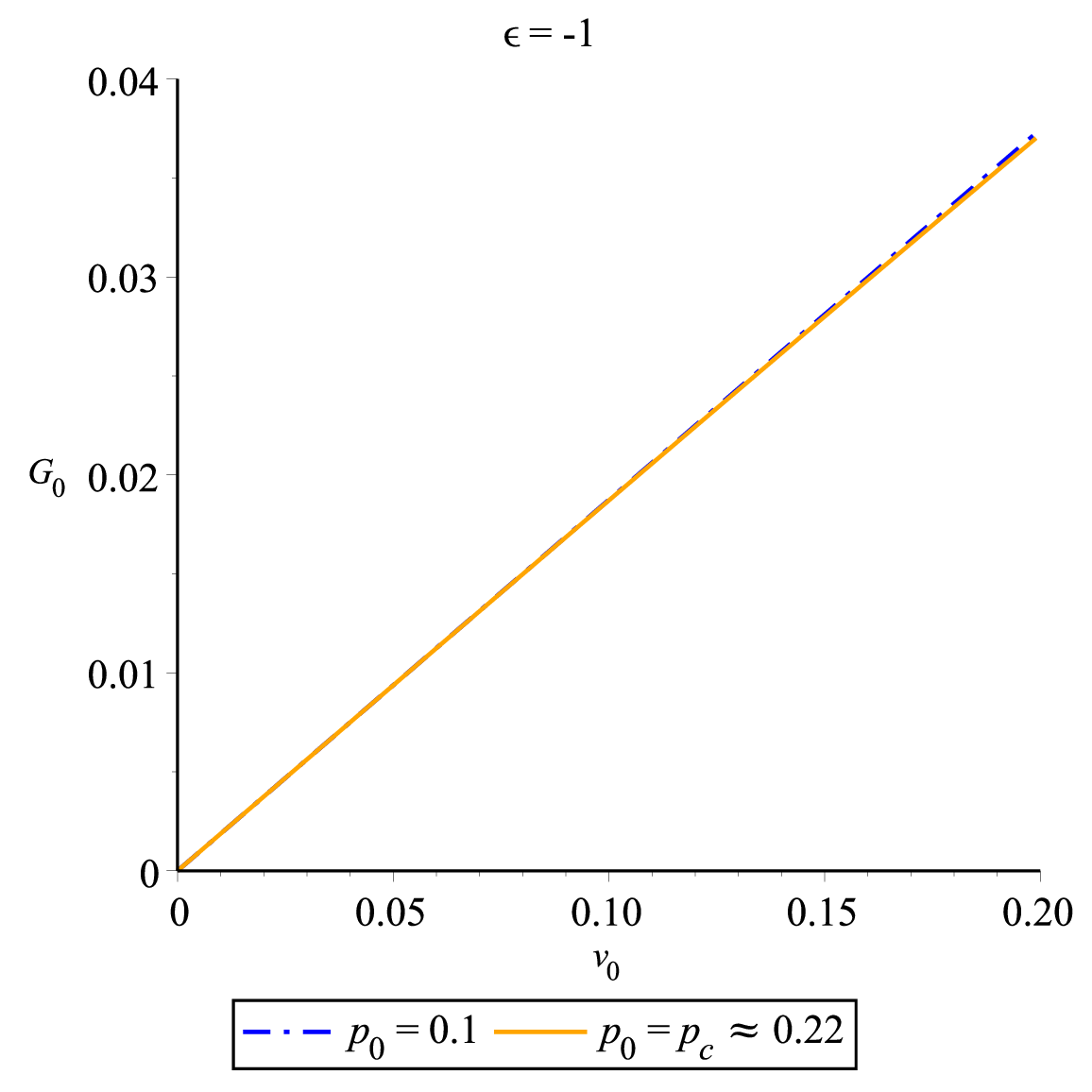}
\includegraphics[width=6cm]{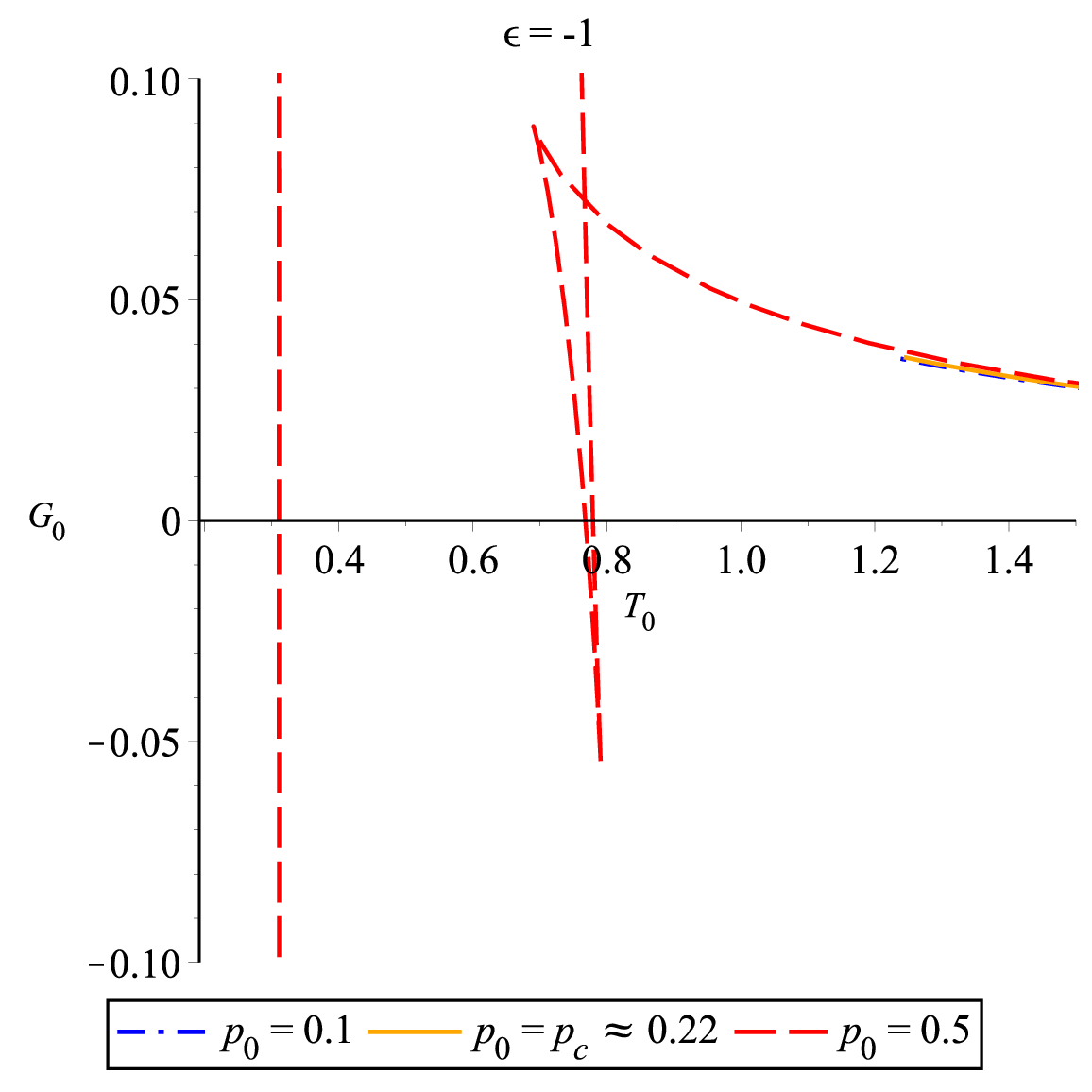}
}
 \caption{Diagrams for Gibbs free energy at ground state n=0 .}
\end{figure}

\begin{figure}[tbp]
\centering{
\includegraphics[width=6cm]{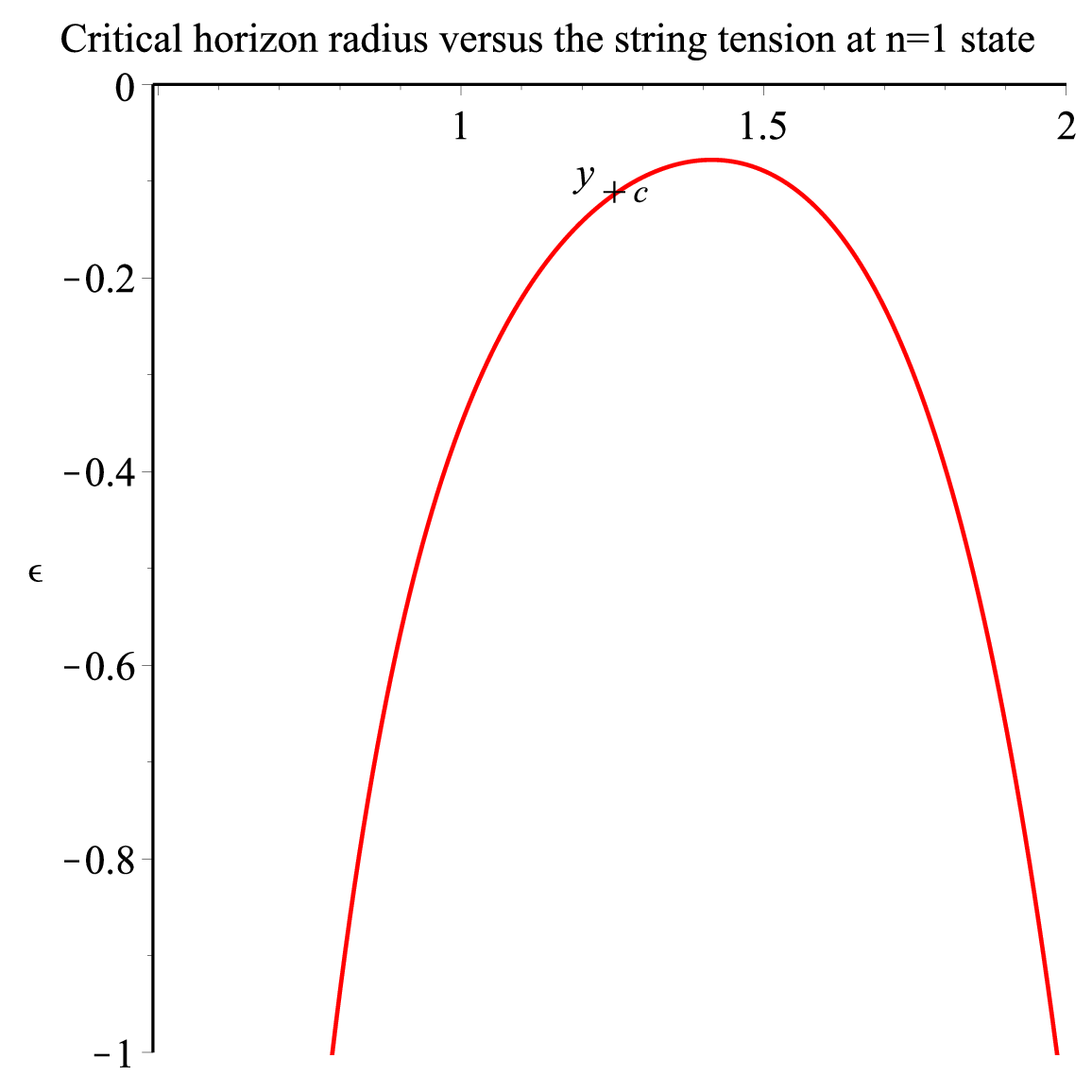}
\includegraphics[width=6cm]{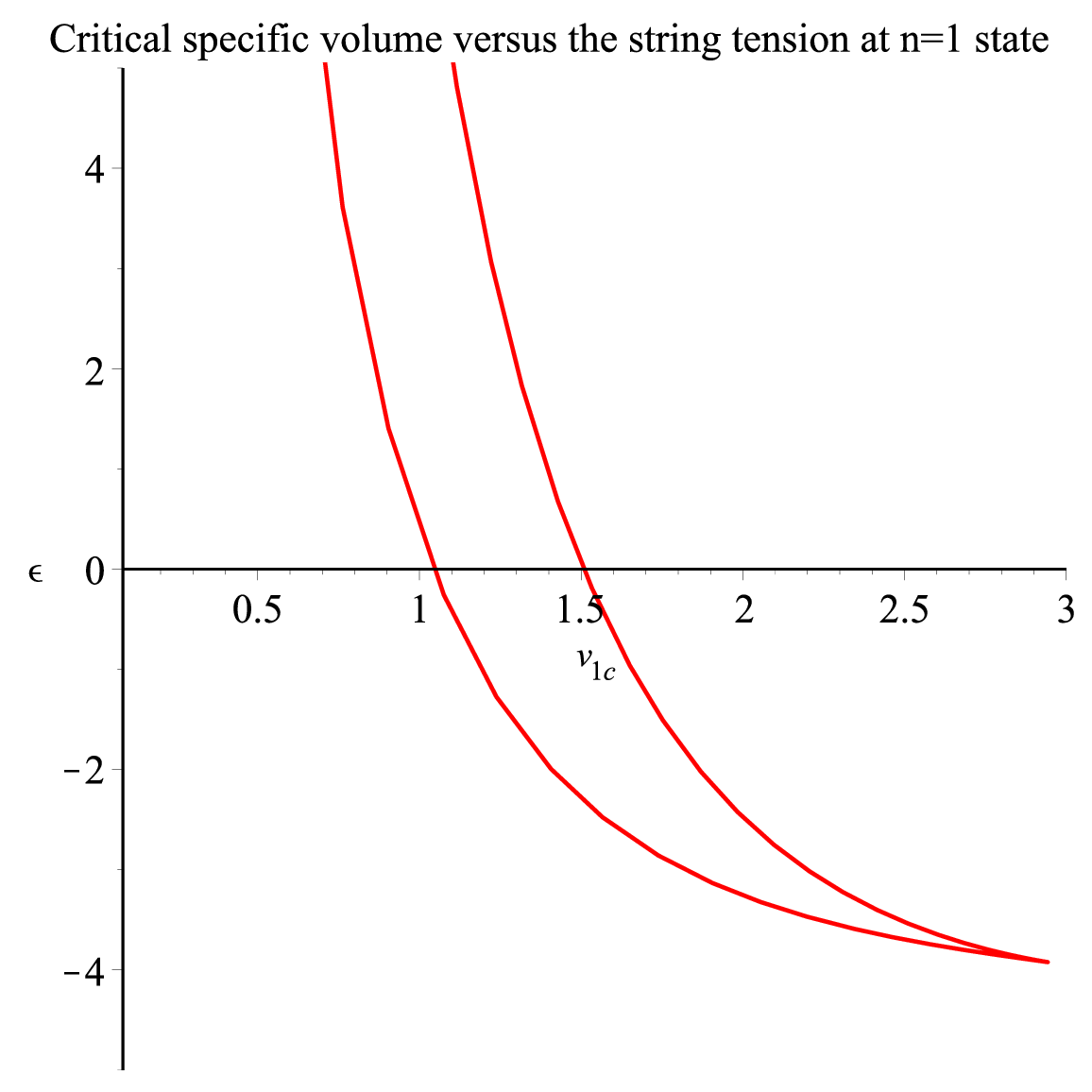}
\includegraphics[width=6cm]{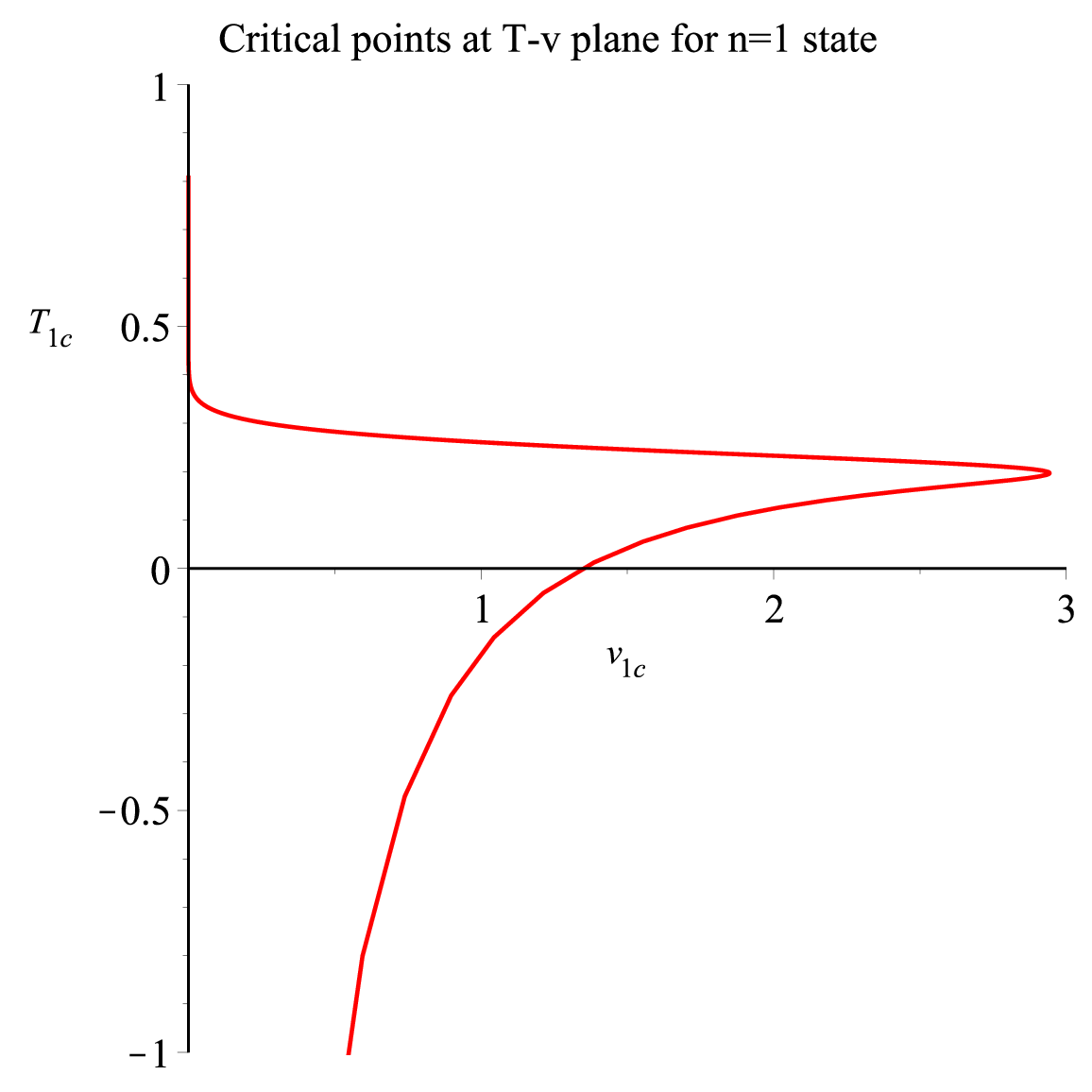}
\includegraphics[width=6cm]{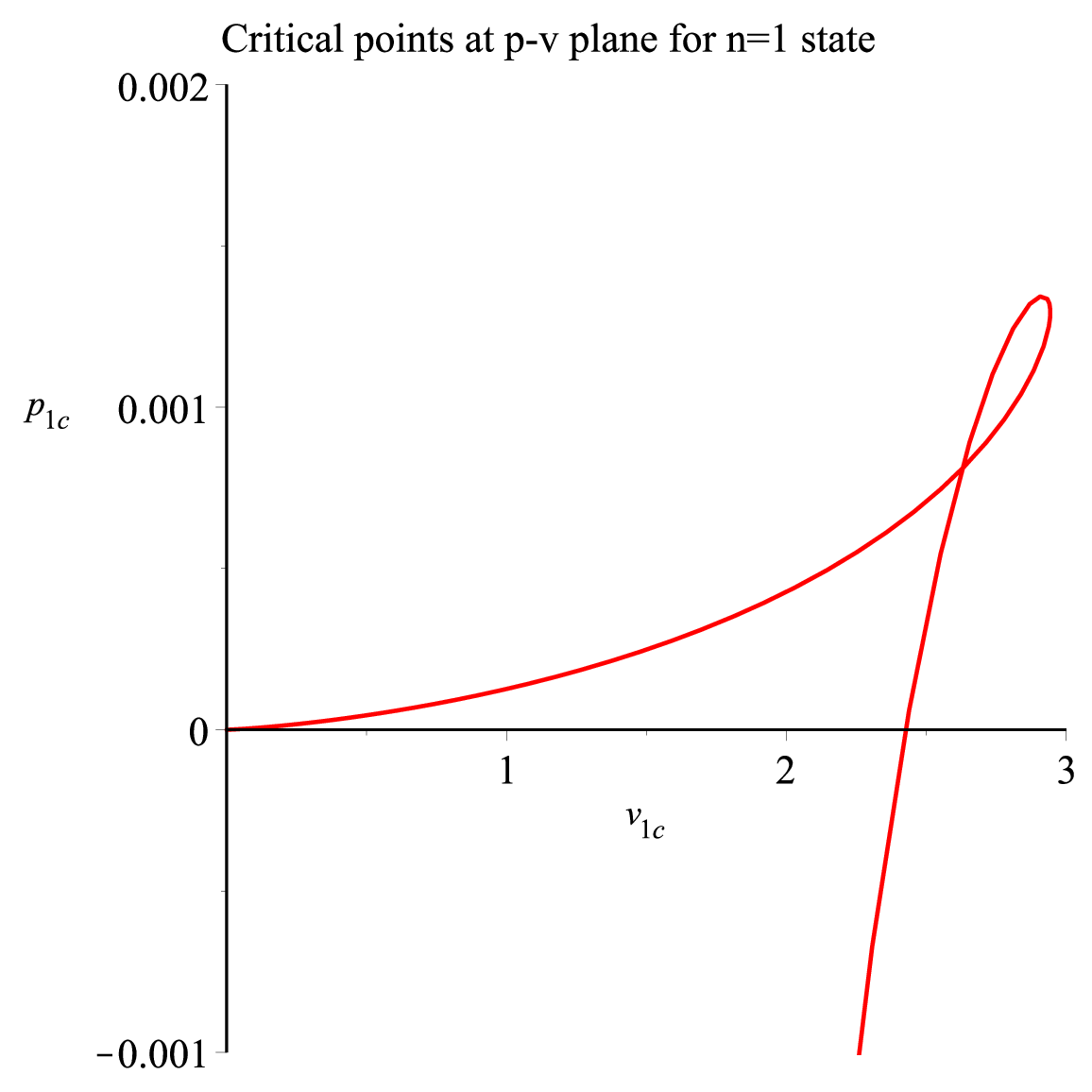}
\includegraphics[width=6cm]{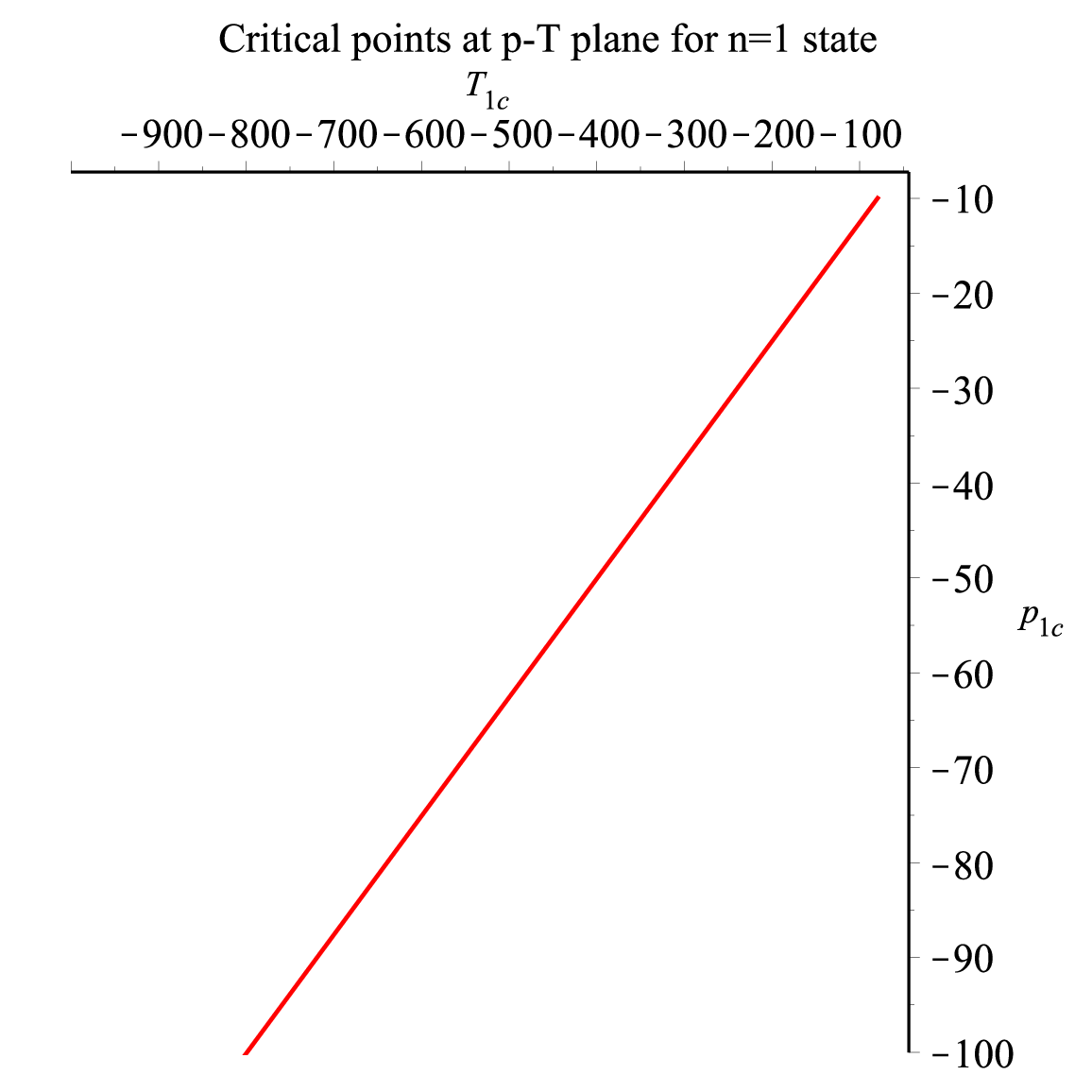}
\includegraphics[width=6cm]{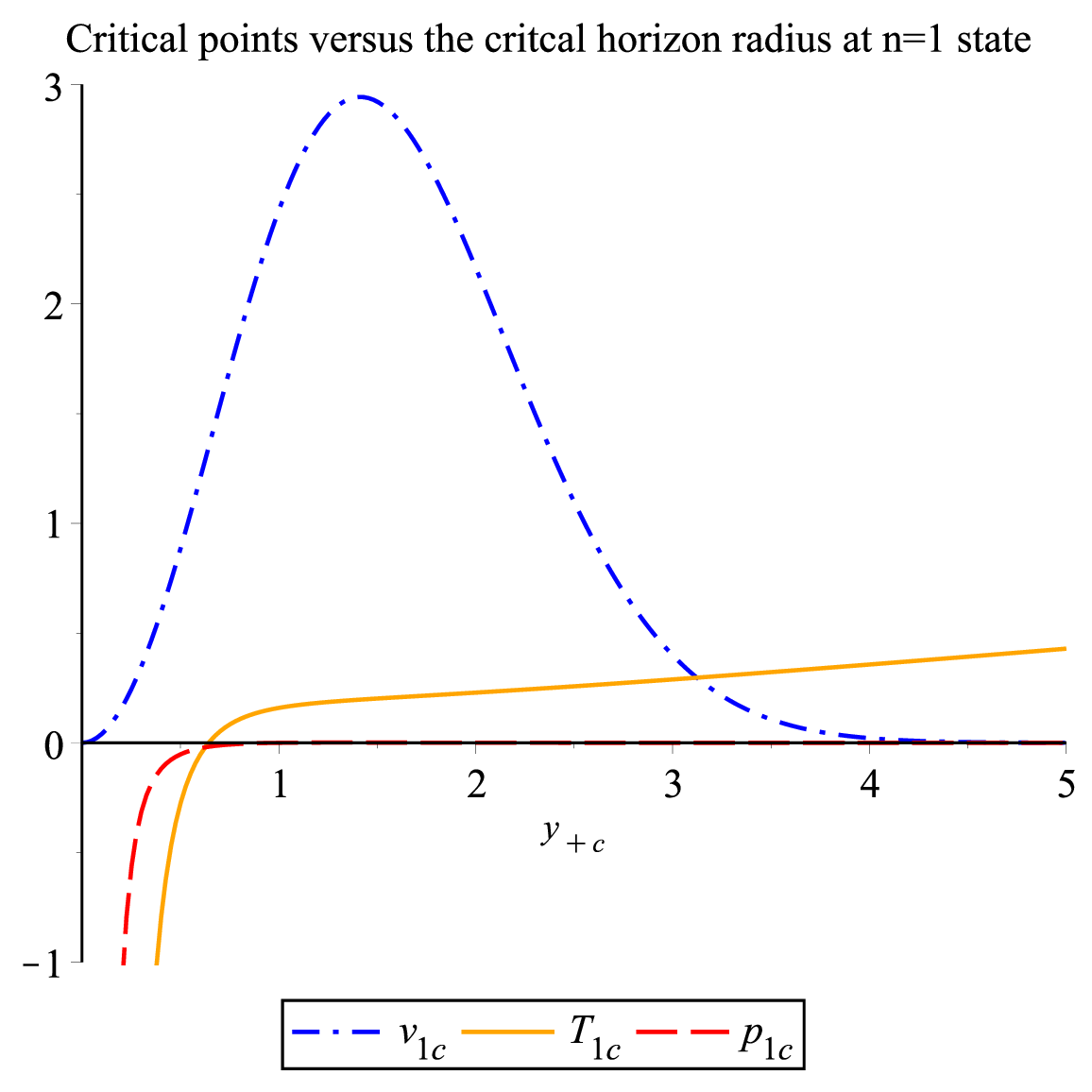}
} \caption{Critical points versus the string tension at first
excited state $n=1$.}
\end{figure}
\begin{figure}[tbp]
\centering{
 \includegraphics[width=6cm]{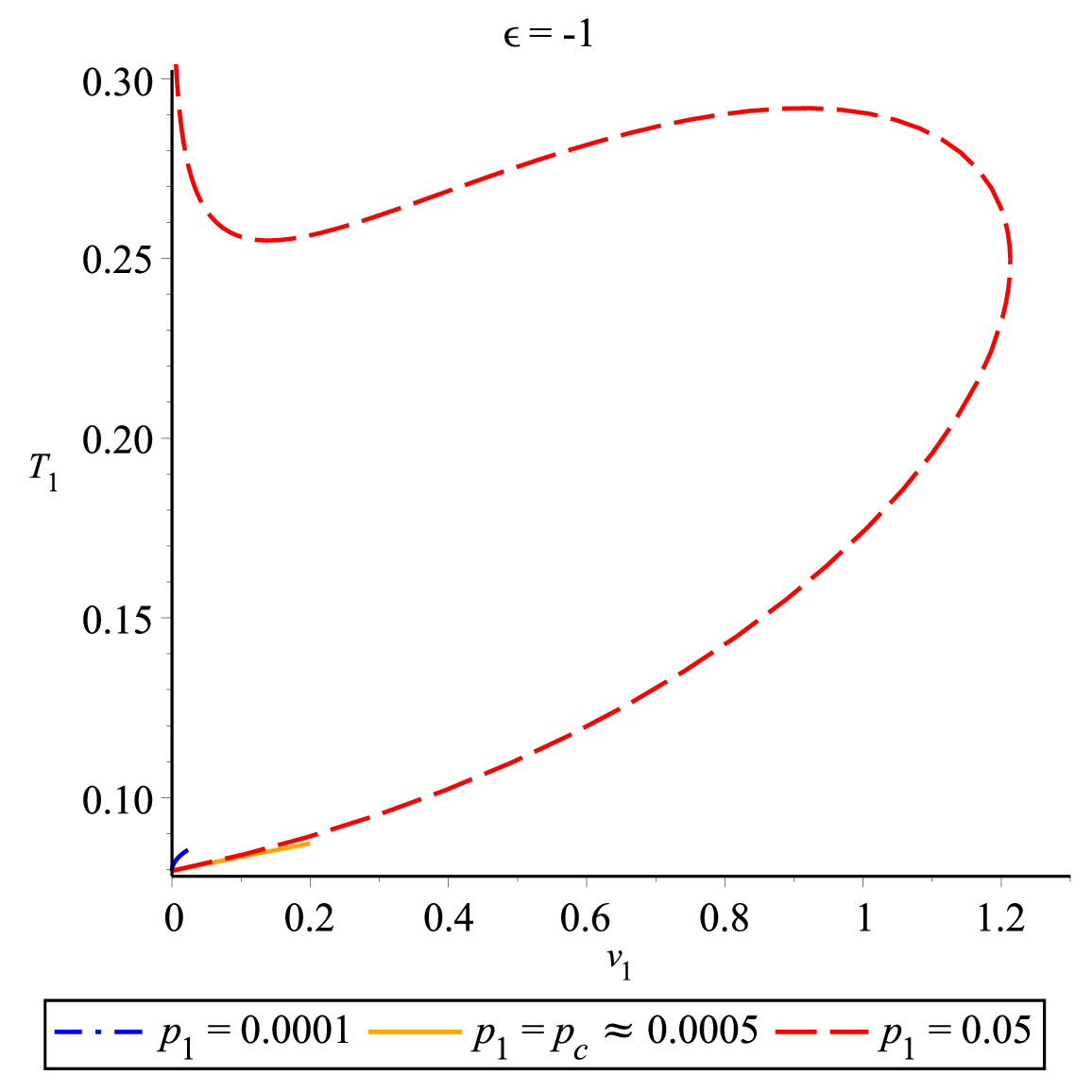}
\includegraphics[width=6cm]{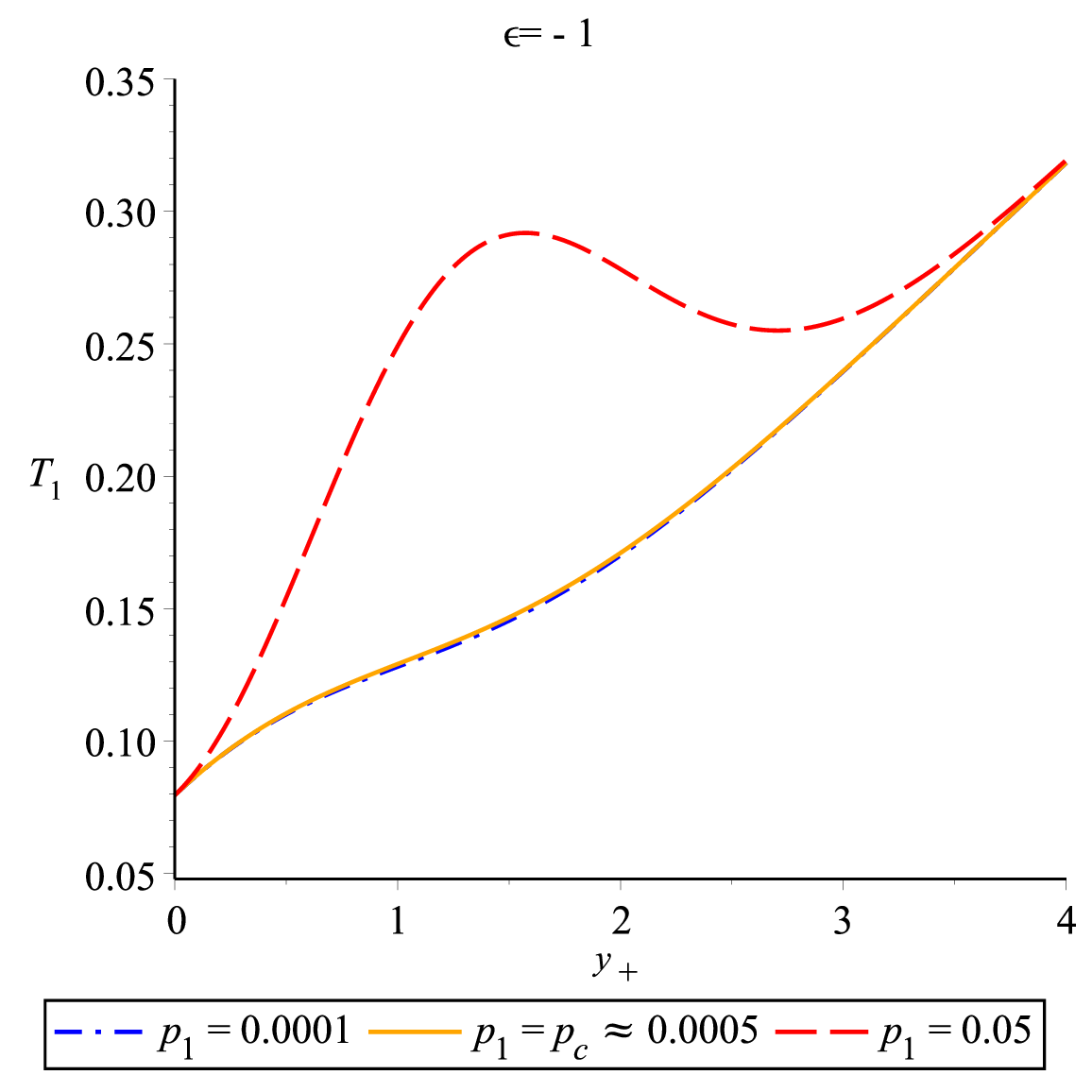}
}
 \caption{Diagram of the temperature is plotted versus the specific volume and the event horizon at $n=1$ state. }
\end{figure}

\begin{figure}[tbp]
\centering{
\includegraphics[width=6cm]{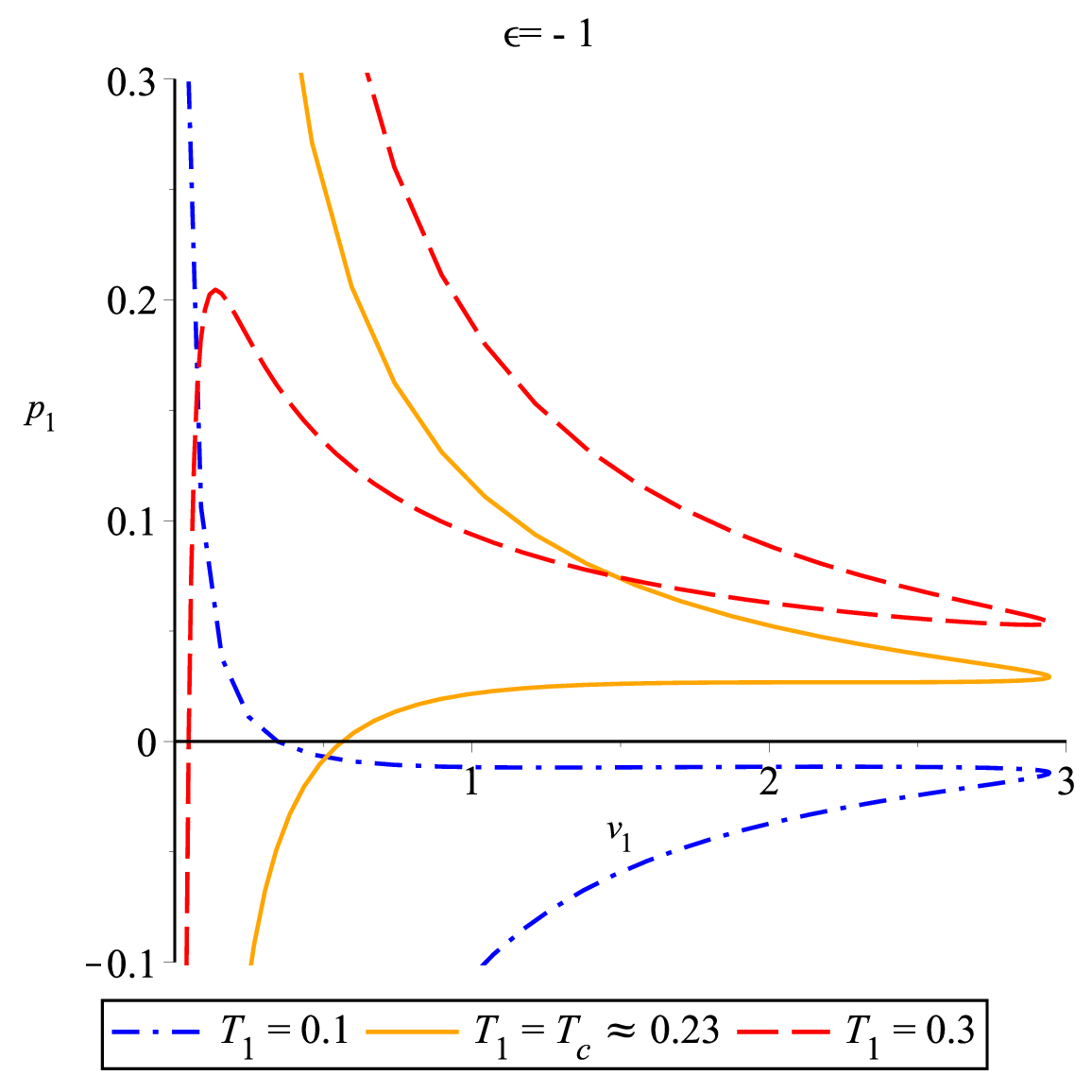}
\includegraphics[width=6cm]{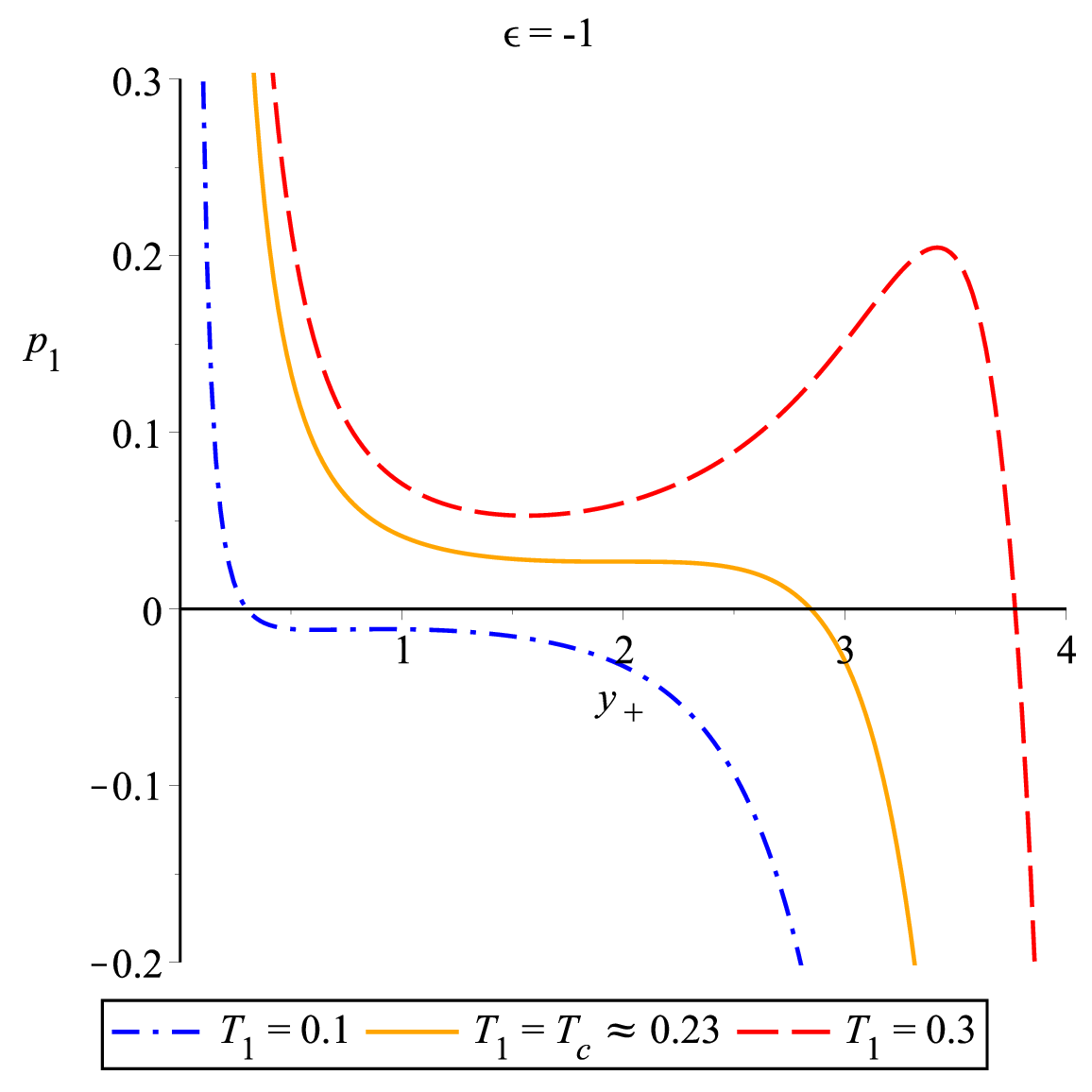}} \caption{Diagram of the  pressure is plotted versus the specific volume and the event horizon at $n=1$ state.  }
\end{figure}

\begin{figure}[tbp]
\centering{
\includegraphics[width=6cm]{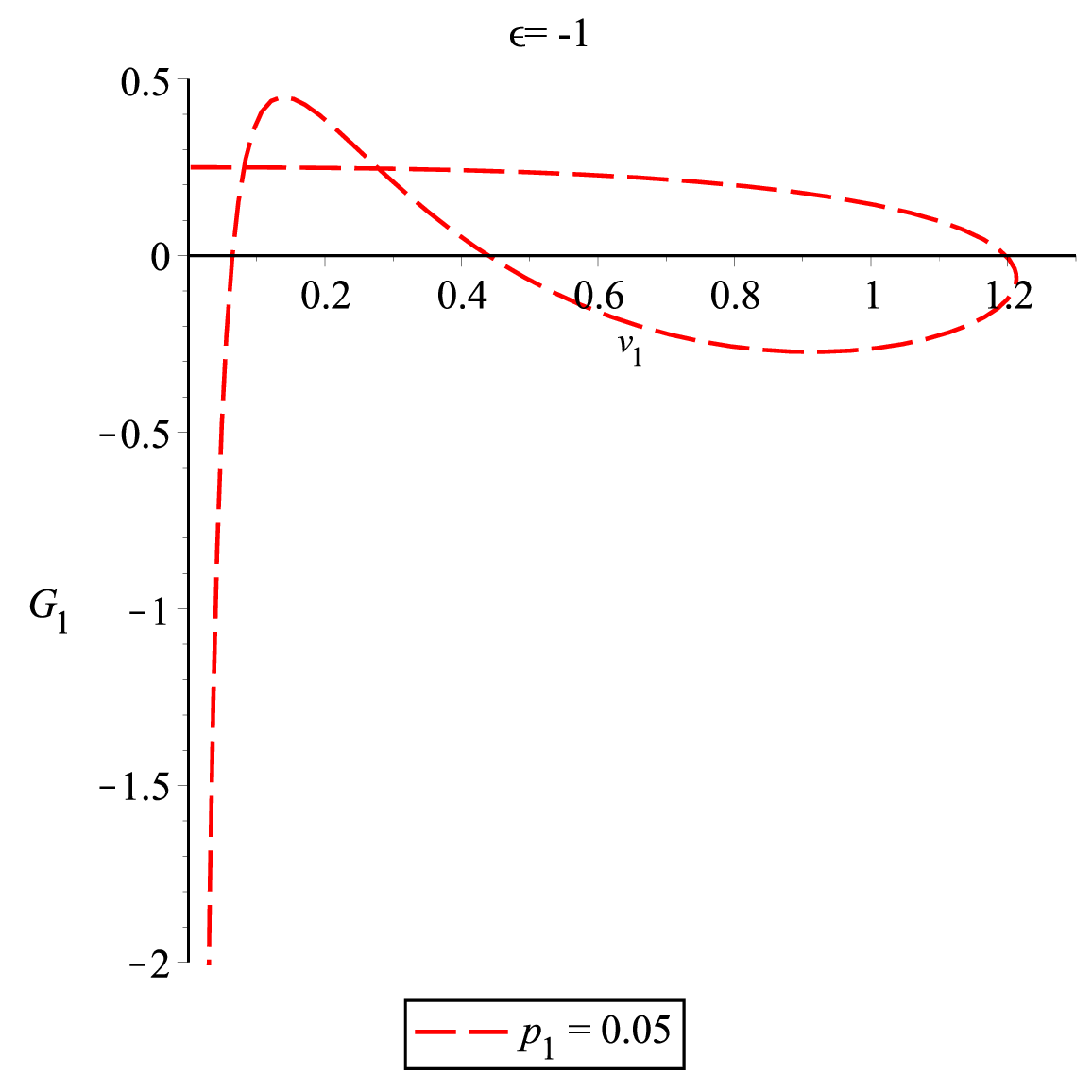}
\includegraphics[width=6cm]{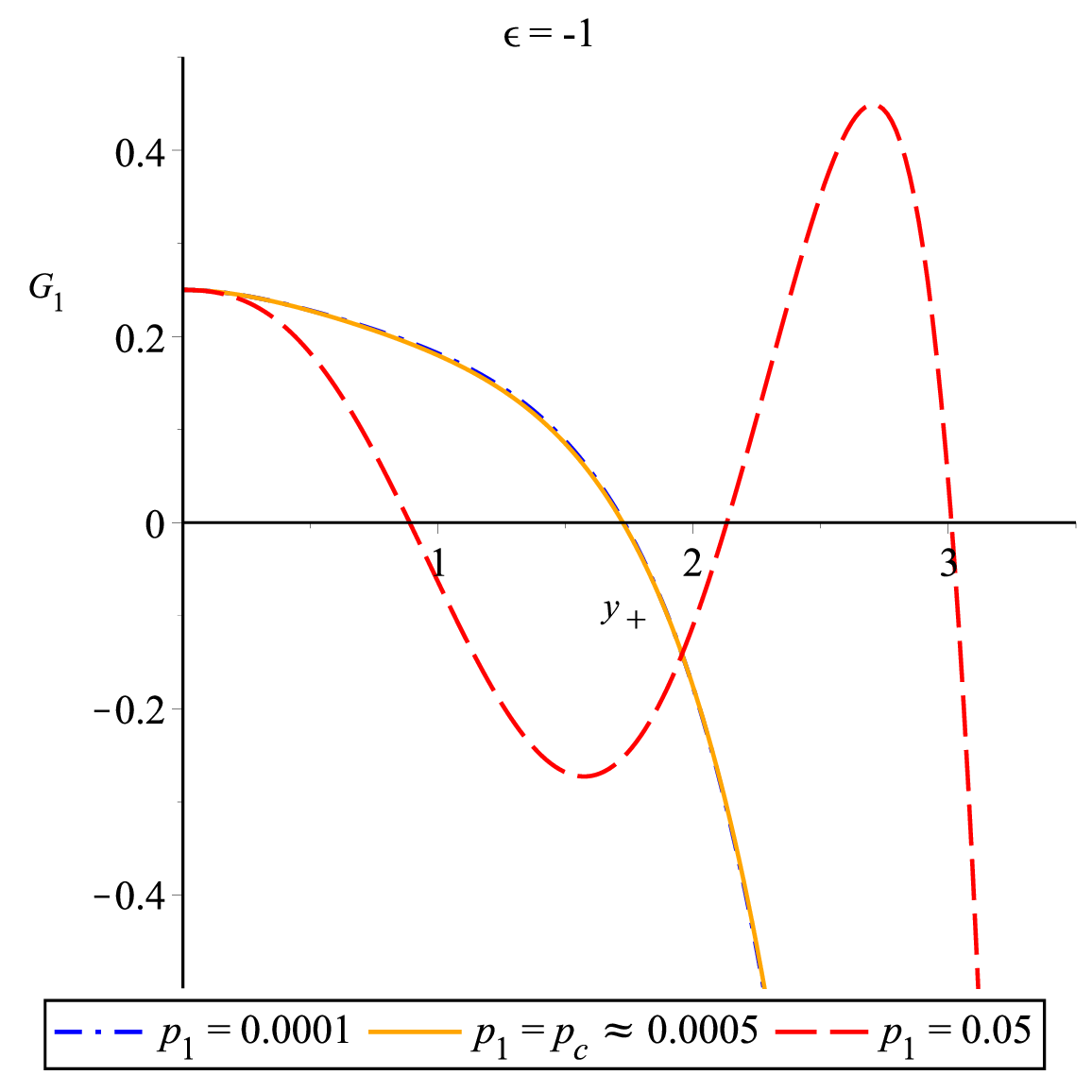}
\includegraphics[width=6cm]{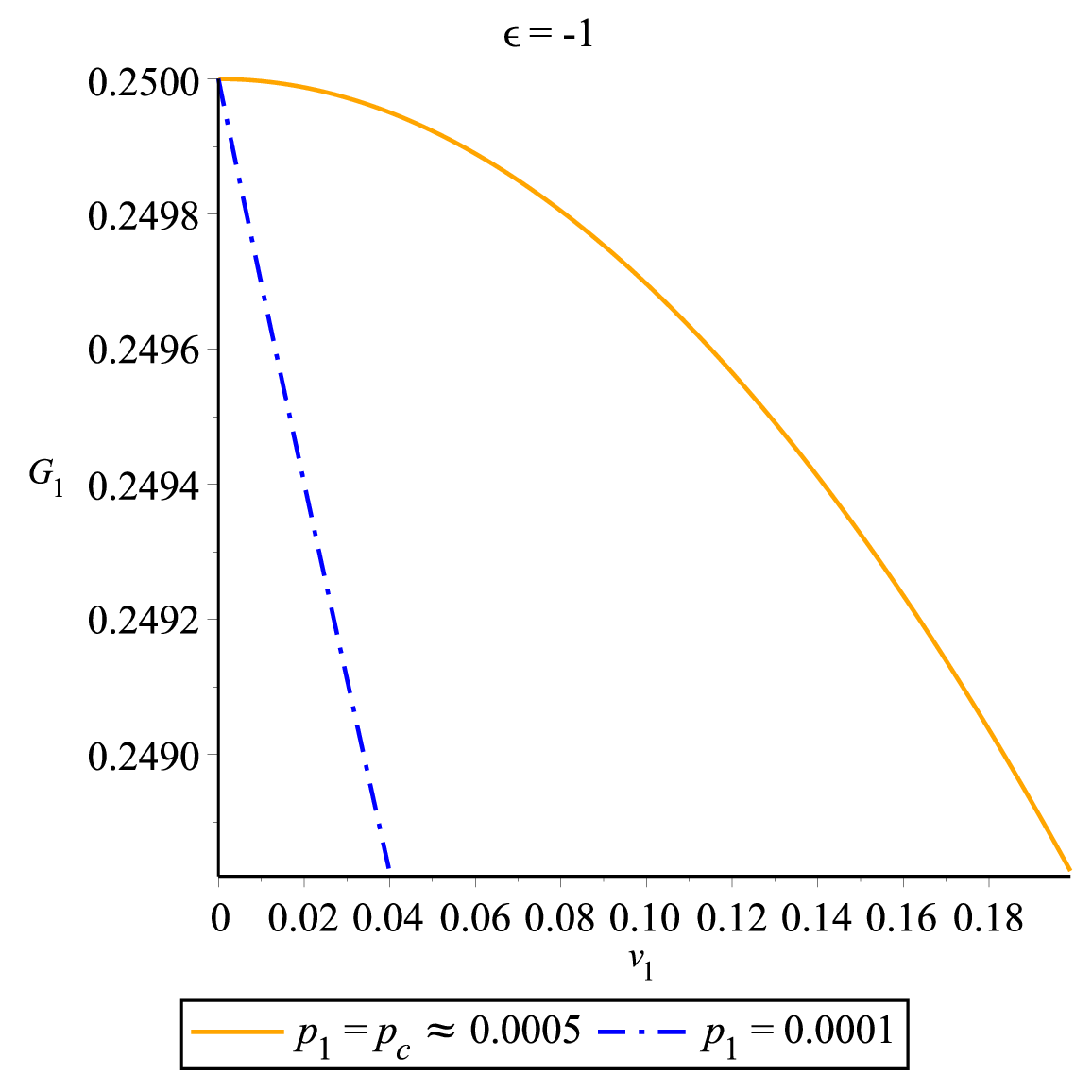}}
\caption{Diagram of the  Gibbs free energy is plotted versus the
specific volume and the event horizon at $n=1$ state. }
\label{Surface_Schw_and_MT}
\end{figure}

\begin{figure}[tbp]
\centering{
\includegraphics[width=6cm]{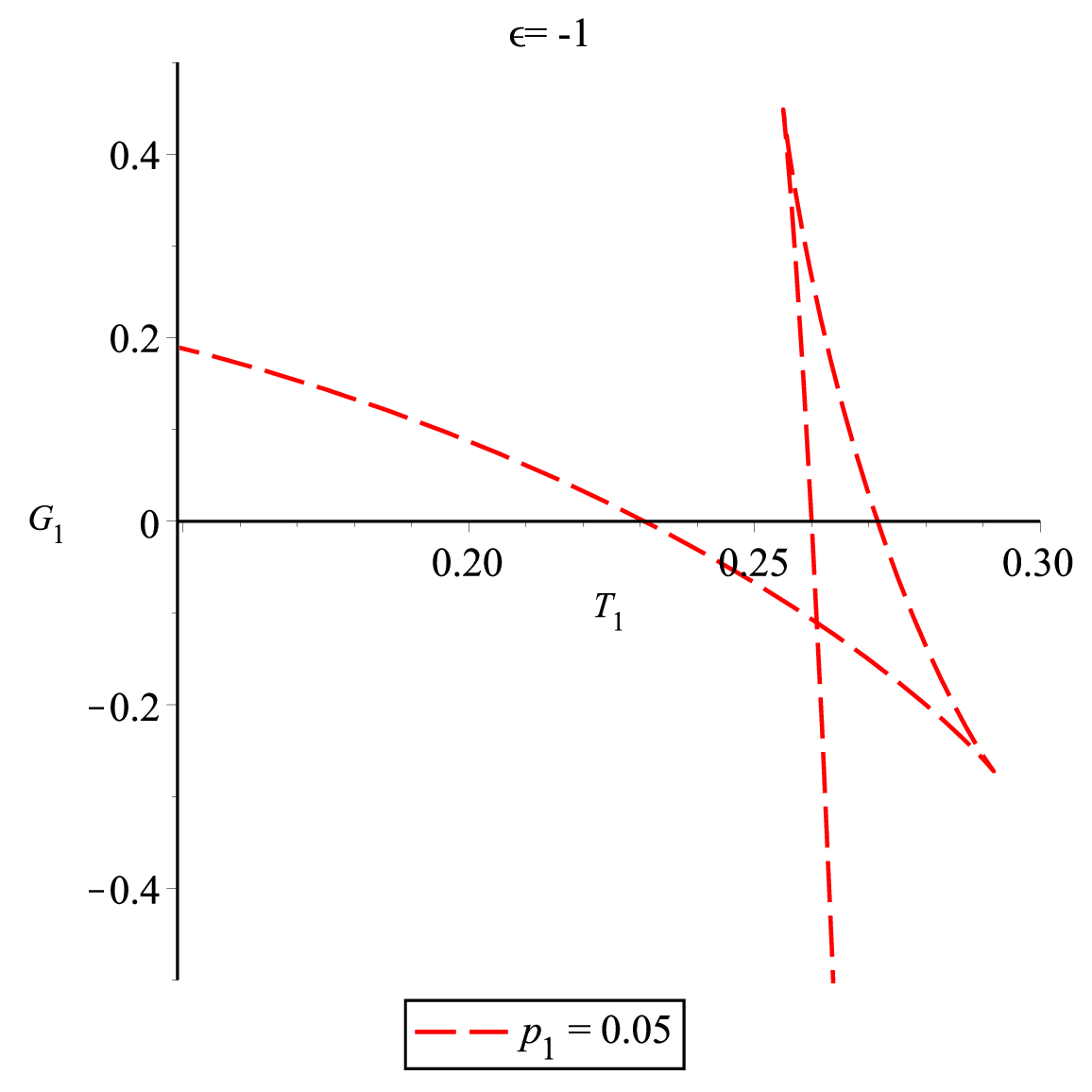}
\includegraphics[width=6cm]{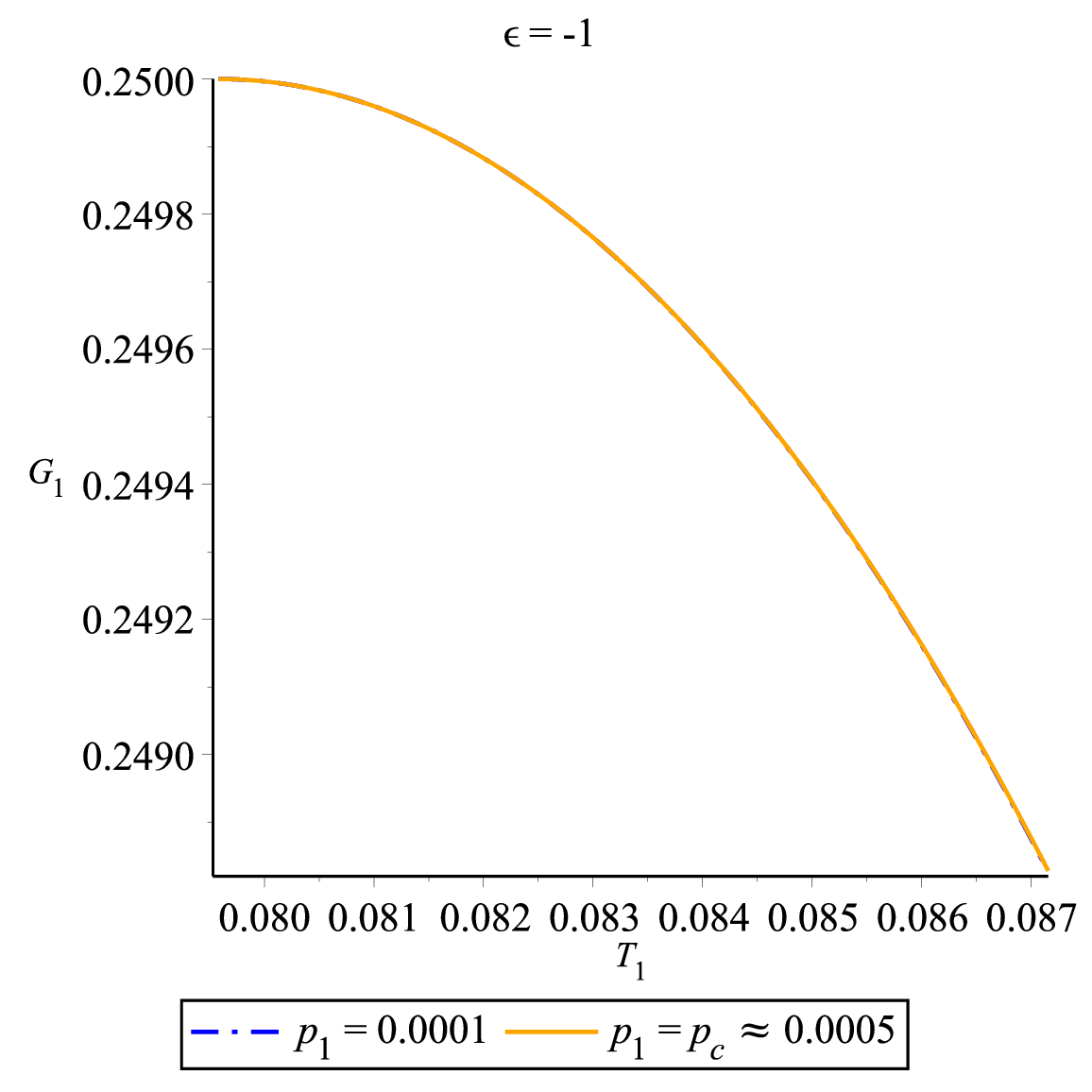}}
\caption{$G-T$ curve is plotted in $n=1$ state for different
values of the constant pressure.} \label{Surface_Schw_and_MT}
\end{figure}

\end{document}